\documentclass[10pt]{article}
\usepackage[top=1in,left=1in,left=1in,right=1in]{geometry} 
\usepackage{indentfirst}
\usepackage[aboveskip=1pt,labelfont=bf,labelsep=period,justification=raggedright,singlelinecheck=off]{caption}
\usepackage{placeins}
\usepackage{soul}
\usepackage{pdflscape}
\usepackage[disable]{todonotes} 
\usepackage{newcent}
\usepackage{amsmath}
\usepackage{subcaption}
\usepackage{booktabs} 
\usepackage{placeins}
\usepackage{makecell}

\usepackage[sorting=none,doi=false, isbn=false]{biblatex}
\bibliography{bibliography}
\AtEveryBibitem{\clearlist{language}} 
\AtEveryBibitem{\clearfield{note}}   
\AtEveryBibitem{\clearfield{urldate}}   
\AtEveryBibitem{\clearfield{urlmonth}}   
\AtEveryBibitem{\clearfield{urlyear}}   
\renewbibmacro{in:}{}
\DeclareFieldFormat{url}{\url{#1}} 
\title{\vspace{-2cm} Predictive modeling of movements of refugees and internally displaced people: Towards a computational framework}
\author{Katherine Hoffmann Pham \\ UN Global Pulse \and Miguel Luengo-Oroz \\ UN Global Pulse}           
\begin{document}
\maketitle
\vspace{-1cm}
\begin{abstract}
Predicting forced displacement is an important undertaking of many humanitarian aid agencies, which must anticipate flows in advance in order to provide vulnerable refugees and Internally Displaced Persons (IDPs) with shelter, food, and medical care. While there is a growing interest in using  machine learning to better anticipate future arrivals, there is little standardized knowledge on how to predict refugee and IDP flows in practice. Researchers and humanitarian officers are confronted with the need to make decisions about how to structure their datasets and how to fit their problem to predictive analytics approaches, and they must choose from a variety of modeling options. Most of the time, these decisions are made without an understanding of the full range of options that could be considered, and using methodologies that have primarily been applied in different contexts -- and with different goals -- as opportunistic references. In this work, we attempt to facilitate a more comprehensive understanding of this emerging field of research by providing a systematic model-agnostic framework, adapted to the use of big data sources,  for structuring the prediction problem. As we do so, we highlight existing work on predicting refugee and IDP flows. We also draw on our own experience building models to predict forced displacement in Somalia, in order to illustrate the choices facing modelers and point to open research questions that may be used to guide future work.  
\end{abstract}

\section{Introduction}
Humanitarian displacements are growing in size and severity. The United Nations refugee agency (UNHCR) estimates that there were 80 million forcibly displaced persons worldwide in 2019, which reflects an increase of over 36 million displaced persons since 2009; displaced persons also represent a growing share of the world's population \cite{unhcr_global_2019, unhcr_unhcr_2021}.  The COVID-19 pandemic has increased displacement pressures and by the end of 2021, it is estimated that there will be more than 95 million {persons} of concern \cite{unhcr_unhcr_2021}.\footnote{ {Formally, a person of concern to UNHCR is ``a person whose protection and assistance needs are of interest to UNHCR. This includes refugees, asylum-seekers, stateless people, internally displaced people and returnees'' \cite{unhcr_glossary_2021}. Additional definitions are given in Appendix \ref{sec:vocabulary}.}} In the face of such massive dislocations, humanitarian agencies have a mandate to monitor the situation and coordinate a response. National governments also require an understanding of these flows for political, {administrative,} and welfare planning purposes.

In order to provide refugees and Internally Displaced Persons (IDPs) with shelter, food, and medical care, these organizations must anticipate flows in advance.  Therefore, predicting forced displacement is an important undertaking of many humanitarian aid agencies, and in practice, such predictions are constantly being made as field offices plan for coming months and request budgets and supplies. More recently, there has been a growing interest in whether computational tools and predictive analytics -- including techniques from machine learning, artificial intelligence, simulations, and statistical forecasting -- can be used to support field staff by predicting future arrivals.

Prior work has adopted a mix of modeling techniques; examined a variety of different refugee, IDP, and migration settings; and attempted to forecast arrivals with different levels of geographic and temporal granularity. For example, \textcite{martineau_red_2010} fits a logistic regression model to estimate refugee generation at the country level. He uses a range of different predictors including measures of conflict, violence, political changes, political freedom, GDP, economic growth, and population size. This model is fairly general, since it predicts only a binary outcome at an aggregate level (whether or not a country generated over 5,000 refugees between 1996 and 2005), but one advantage of his approach is that it sheds insight into possible structural drivers of refugee {movements} in the long term.

In contrast, \textcite{suleimenova_generalized_2017} use Agent-Based Models (ABMs) to predict the distribution of refugees across available camps in three emergency settings in Burundi, the Central African Republic (CAR), and Northern Mali. They model individual behavior in a detailed environment, which is implemented as a graph connecting conflict regions, transit hubs, refugee camps, and refugee settlements. At each time period, refugees are displaced and they select a destination according to the attractiveness of that destination. They also face a cost to travel that is based on the real difficulty of traversing the local road network. Using a simulation that advances daily, the authors attempt to replicate the observed patterns of arrivals across refugee camps. On the other hand, \textcite{sokolowski_modeling_2014} propose an even more specific ABM focused on population displacement in Aleppo, Syria. Their model consists of three types of actors -- government agents, rebels, and civilians. Government agents are motivated by social considerations, emotion, and rational self-interest to conduct spatial attacks on rebels operating within a gridded model of the city. As a result of these attacks, civilians are {displaced}, and the authors use the model to predict the fraction of the city's population that will eventually be displaced.

More recently, increased access to data and advances in computational methods \cite{drouhot_computational_2022} have facilitated the construction of machine learning models for predicting displacement. For example, in the context of internal displacement, \textcite{huynh_forecasting_2020} attempt to forecast pairwise IDP flows between provinces in Syria and Yemen, using variables including conflict, prices, wages, and distances. They test a range of statistical and machine learning models, including: {linear mixed effects models, support vector machines (SVMs), random forests, XGBoost, and multi-layer perceptrons.} On the other hand, for the more general context of international mixed migration, \textcite{nair_machine_2020} attempt to predict pairwise flows of migrants exiting Ethiopia for six destination countries. They use indicators of a wide range of migration drivers ranging from health infrastructure to unemployment rates, and test models such as linear regressions, random forests, SVMs, and XGBoost.

While a number of organizations have begun to pilot such predictive analytics projects,\footnote{~The UN Office for the Coordination of Humanitarian Affairs (OCHA) maintains an ongoing list of {predictive analytics} projects at: \url{https://centre.humdata.org/catalogue-for-predictive-models-in-the-humanitarian-sector/}. {A helpful review of different predictive analytics initiatives in the humanitarian sector can be found in \textcite{hernandez_predictive_2020}.}} there is a lack of knowledge and standardized protocols for predicting refugee and IDP flows in practice. Practitioners are confronted with a variety of decisions about how to structure their datasets and how to fit their problems to standard predictive analytics approaches, and they must choose from a variety of competing modeling approaches without having a comprehensive understanding of the potential options and their suitability. Despite the need for guidance on these choices, academic work on the subject is limited. 

 In this article, we attempt to address this gap by providing a computational framework for leveraging big data sources to create predictive models of forced displacement. We seek to provide guidance from a practical standpoint, discussing the tradeoffs modelers face when framing problems in this context. 
 In Section \ref{sec:problem_def}, we define the types of problems covered by our framework. In Section \ref{sec:data}, we discuss the data sources available for the analysis of forced displacement. In Section \ref{sec:decisions}, we discuss the decisions involved in defining a problem statement and applying this data to the prediction of humanitarian flows. In Section \ref{sec:implementation} we discuss implementation, reviewing the types of models that are available for prediction as well as some of the practical concerns involved with deploying these models in context. To enhance this discussion, we draw on examples from previous work on predicting refugee and IDP displacement, as well as a case study from our own experience forecasting IDP flows in Somalia, which we describe in Section \ref{sec:somalia}. 
Finally, in Section \ref{sec:discussion} we outline a set of open research questions that represent promising directions for future work. {To complement this article, we have also produced a practitioner-oriented visualization of the framework, which we describe further in Appendix \ref{sec:modeling_cards}.} 

Writing in 1998, \textcite{schmeidl_early_1998} called for researchers to ``develop new modeling and data development techniques'' to improve early warning systems; ten years later, \textcite{edwards_computational_2008} lamented the fact that ``there has been little systematic attempt to use computational tools to create a practical model of displacement for field use.'' In the intervening ten years the range of datasets and modeling techniques available to researchers has grown significantly, but in practice little has changed. Our hope is that we will facilitate a more comprehensive understanding of this emerging field of research by providing a systematic, model-agnostic framework that is able to leverage multiple data sources for structuring the prediction problem. As \textcite{suleimenova_generalized_2017} note, such general model development frameworks can be more useful than well-calibrated but situation-specific models, because new models must be developed rapidly when unexpected crisis situations emerge. 
We hope that this article provides guidance in support of the renewed interest in predicting displacement, and provides the practical background to facilitate a long-overdue advancement in the state of the art.

\section{A computational framework for predicting forced displacement}\label{sec:problem_def}
Humanitarian teams typically face three types of modeling needs: \textit{description} of current or previous situations; \textit{prediction} of what will happen in the future; and \textit{simulation} of possible alternative scenarios. We focus on prediction, although these three needs are not entirely distinct: conducting descriptive analysis to understand a given context is often a necessary prerequisite to building well-informed predictive models, and simulations can also be used to generate predictions. A more detailed discussion of these modeling needs can be found in {Appendix} \ref{sec:modeling_needs}.

Our framework focuses on modeling the movement of refugees and IDPs, rather than migrants more generally; formal definitions of these terms are provided in {Appendix}  \ref{sec:vocabulary}. In principle, we take a flexible approach to defining the specific population of interest, since this will vary according to the modeler's needs and the available data; the discussion below should apply to most involuntary or emergency displacements. Complementary approaches include modeling {international migration processes, a field with a rich literature \cite{massey_theories_1993, sirbu_human_2020,  bijak_forecasting_2006, sohst_forecasting_2021, molina_how_2022}}. 

The term ``displacement'' encompasses a wide variety of movements, which may occur for different reasons and over different time scales (see {Appendix}  \ref{sec:displacement_drivers} for further discussion). The framework below focuses on the challenges posed by studying:
		\begin{itemize}
			\item Short-horizon, weeks- and months-long (rather than years- or decades-long) movements;
			\item Rapidly-changing and unexpected (rather than gradual) displacement processes;
			\item Emergent, data-scarce (rather than well-studied) contexts;
			\item Local, sub-national, or site-specific (rather than country-level) populations.
		\end{itemize}	 

We approach the discussion from a perspective informed by data science, machine learning, and engineering approaches. This is a notable departure from much of the early work on forced displacement, which tended to focus on a social science perspective. 
One consequence of adopting a data-driven approach is a greater tolerance for the use of features that are predictive, but not necessarily informative. As an example of such a feature, consider the price of a hired taxi between two cities; this price might be closely correlated with the magnitude of displacement flows between the cities, but {reveals} relatively little about the underlying drivers or the nature of the displacement. In contrast, much of the previous research on migration and displacement has prioritized an understanding of the structural drivers and mechanisms of displacement, often by choosing interpretable models (e.g., linear regressions) at the expense of predictive accuracy. While we do not weigh in favor of one approach or another (and in fact believe that the strongest approaches combine both perspectives), we feel that the data science and machine learning perspective is much less prevalent in the field and therefore deserves serious consideration from researchers in the future.

The computational framework described below is composed of three phases: (1) data collection; (2) the definition of the prediction problem and the associated decisions taken by the modeler; and finally, (3) model implementation.  {These phases are not necessarily sequential and multiple iterations might be needed; for example, after defining the prediction problem there may be an interest in gathering additional data, or after implementing a model in practice, there may be a need for improvements which require changes in the original model structure.}

\section{Data on forced displacement}\label{sec:data}

A precursor to the design and development of predictive models is the gathering of relevant data, and improvements in the collection and availability of data in recent years have made it possible both to better capture displacement flows, and to disentangle the drivers and nature of these flows. Modelers today can choose from a wide variety of traditional and non-traditional data sources, which we describe in more detail below.

\subsection{Displacement data}\label{sec:prmn}

While there is no comprehensive source of data on refugee movements and internal displacement, in our experience the most helpful indicators are collected by UNHCR {and} the International Organization for Migration (IOM). In particular, IOM's Displacement Tracking Matrices (DTM) monitor sub-national displacement flows, and typically consist of a series of datasets capturing population movements at different time points \cite{iom_displacement_2020}. The contents and design of DTMs may be country- and situation- specific, but at time of writing 81 countries are covered by various DTM data collection processes. One consideration in using this data is that data collection appears to be driven by the relevance/timeliness of a given displacement emergency, so the amount and frequency of historical data varies by country. For example, Iraq has undergone 120 rounds of measurement as of March 2021, whereas the Philippines have undergone only 5 rounds that were specifically focused on natural disasters. 

In addition to these subnational flows, the IOM curates a helpful dataset on individual-level deaths in the course of migration \cite{iom_missing_2021}. UNHCR also maintains an operational portal with situation-specific data on displacement emergencies \cite{unhcr_operational_2020}. A final source of data is the International Displacement Monitoring Center (IDMC), which maintains a global dataset of estimated displacement counts by country \cite{idmc_global_2019}.  The data are disaggregated according to whether the driver of displacement is conflict- or disaster-related, as well as whether the population in question has been newly displaced or not.

\subsection{Conflict and political data}

A number of different data collection initiatives provide information on conflict, violence, and political repression. One easily accessible source of near real-time data is the Armed Conflict Location and Event Database (ACLED), a global database that provides incident-level data via its API \cite{raleigh_introducing_2010, holmes_acled_2019}. The data includes information on the event location, date, administrative region, actors involved, description, and source. Events are also categorized according to ACLED's own taxonomy \cite{acled_acled_2017}, which consists of: violent events (battles, explosions/remote violence, and violence against civilians), demonstrations (protests and riots), and non-violent actions (strategic developments). 

Alternative sources of conflict {and event} data include the Integrated Crisis Early Warning System (ICEWS) dataset \cite{boschee_icews_2020}; {the GDELT Project \cite{the_gdelt_project_gdelt_2015};} and the Uppsala Conflict Data Program \cite{uppsala_university_uppsala_2018}, although the latter provides only historical data. Country-level data on political repression can be obtained from Freedom House \cite{freedom_house_freedom_2020} and the Center for Systemic Peace's Polity Score datasets \cite{center_for_systemic_peace_polity_2018}.  A helpful review of conflict data collection efforts and how these efforts have shaped conflict research can be found in \textcite{gleditsch_advances_2020}.

\subsection{Cell phone data}
As more of the world's population relies on mobile communications, one promising means of tracking mobility flows is Call Detail Records (CDR), which consist of information collected by mobile phone companies each time a call is placed. CDRs typically contain information on the the caller and the recipient, the start time and duration of the call, and the cell phone towers that were nearest to the caller and the recipient at the time the call was placed. By tracking an individual's calls over time, it is possible to identify the individual's patterns of movement; using the individual's call log, it is also possible to construct the individual's social network. CDR data was used by \textcite{lu_predictability_2012} to estimate displacement after the Haiti earthquake; the study found  that individual movements were highly predictable, and correlated with the regions to which the individuals were connected according to their call history and past travel patterns. CDR data has also been used to track displacement in Nepal \cite{wilson_rapid_2016}; the impact of flooding in Mexico \cite{pastor-escuredo_flooding_2014}; responses to drought in Colombia \cite{isaacman_modeling_2018}; seasonal migration in Senegal \cite{zufiria_identifying_2018}; and the mobility patterns of Syrian refugees in Turkey \cite{salah_introduction_2019}.

\subsection{Remote sensing and geographic data} 

There has also been a growing interest in the use of remote sensing  for tracking humanitarian displacement. For example, researchers have used nighttime lights data -- which essentially consists of pixellated snapshots of a geographic area in which each pixel reflects light intensity -- to monitor the development of the conflict in Syria \cite{li_can_2014}. A richer source of data is satellite images, which can show more detailed footprints of conflict, evidence of displacement-inducing environmental stressors such as drought \cite{michella_sentinel-2_2019}, and the emergence of human settlements. In addition to manually labeled satellite images, recent developments in artificial intelligence -- and specifically, Convolutional Neural Networks (CNNs, \cite{lecun_deep_2015}), which are neural networks that are particularly well suited to identifying spatial patterns in images -- have facilitated the automatic labeling of images at scale. Ongoing efforts are studying the transferability of these algorithms across different displacement scenarios, and experimenting with human-in-the-loop approaches to training these algorithms \cite{logar_pulsesatellite_2020, quinn_humanitarian_2018}.  

Finally, mapping platforms are also able to provide relevant geographic information. For example, commercial map providers such as Google \cite{google_developers_google_2020} can generate driving directions and travel time estimates for different transportation modes given an origin-destination pair. The crowdsourcing platform Open Street Map also offers a routing API that makes it possible to freely obtain driving directions for a given origin-destination pair \cite{open_street_map_overpass_2020}.

\subsection{Economic and environmental data}
Other variables of interest  include economic and environmental variables. {For example, in Somalia we incorporated data from the Food Security and Nutrition Analysis Unit (FSNAU) of the United Nations Food and Agriculture Organization (FAO), which maintains a comprehensive database of average commodity prices by market and month} \cite{fsnau_integrated_2019}, as well as a dashboard early warning system which contains data on rainfall, river levels, vegetation cover, and illness \cite{fsnau_early_2019}. Candidate indicators specific to individual displacement settings can often be found on OCHA's Humanitarian Data Exchange, which is a platform that allows humanitarian agencies to share datasets in standardized formats \cite{ocha_humanitarian_2020}. 

\subsection{Social media data}
Other promising big data sources include social media platforms such as Facebook and Twitter. On the one hand, text analysis of online social media discourse with the help of Natural Language Processing (NLP) tools can be used to construct measures of local sentiment, capturing factors such as social unrest, political instability, xenophobia, hate speech, and public perceptions of the government. For example, an analysis of Twitter discussions enabled UN Global Pulse and UNHCR to track sentiment towards migrants along popular transit routes through Europe at the height of Syrian refugee crisis \cite{un_global_pulse_and_unhcr_innovation_service_social_2017}. On the other hand, social media platforms also generate digital trace data that can be used to estimate population movements. For example, Facebook audience estimates (i.e., the platform's estimates of the number of users in different demographic groups) have been used to estimate population displacement and subsequent returns to Puerto Rico after Hurricane Maria \cite{alexander_impact_2019}, and the distribution of Venezuelans living abroad \cite{palotti_monitoring_2020}. Google search query volumes have even been used as a measure of international migration intentions \cite{golenvaux_lstm_2020}.

%%%%%%%%%%%%%%%%%%%%%%%%%%%%%%%%%%%%%%%%%%%%%%%%%%%%%%%%%%%%%%%%%%
\begin{table}[t]
 \centerline{
\begin{tabular}{|l|l|}
\hline
\textbf{Decision} & \textbf{Key questions to ask} \\
\hline
 Unit of Analysis & What is the population of interest -- migrants, refugees, IDPs, all? \\
						 & Does the model capture the behavior of individuals or aggregate populations? \\
						 & At what geographic level does aggregation occur -- national, administrative units, cities, camps?\\
\hline
Time horizon & Is the data longitudinal (panel) or cross-sectional?  \\
						 & What is the frequency of observation -- daily, weekly, monthly, quarterly, annually? \\
						& How much historical data will be incorporated? \\
                & What is the forecast horizon -- how far into the future will the model predict?  \\
 \hline
 Target variable & Does the target variable consist of inflows, outflows, internal flows, or pairwise flows? \\ 
                & Is the target variable binary (classification) or continuous (regression)? \\
				& Will the target variable be transformed? \\
				& Will the model predict the level of the indicator, changes in the indicator, or categorical alerts? \\
\hline
Feature variables & Which contextual features will be included -- environment, health, conflict, governance? \\
								& Are these features available in a standardized manner across all regions? \\
                                    & Will the model include historical lags of the target or feature variables? \\
									& Will predictions use data from nearby (or all) regions? \\
									& What approaches will be used for feature selection, {normalization}, and regularization? \\
\hline
Missing data,  	
   & How will different datasets be integrated and standardized? \\
  data quality, & {What are the uncertainties in data collection?} \\
 and data scarcity & By what processes does missing data arise? \\
		 &  Do these processes bias the dataset or the resulting models? \\
		 & How will the model handle missing data -- dropping, imputation, or one-hot encoding? \\
\hline
Modeling approach
               & What methodology will be used -- econometrics, agent-based modeling, machine learning? \\
\hline	
Model selection & How will cross-validation be implemented? \\ 
			   & {How will the results of multiple models be compared and/or combined?}\\
\hline
Model performance & What error metric will be used to train the model? \\
& What is the appropriate benchmark for model evaluation? \\	
& {How will uncertainties be measured and quantified?} \\		
\hline
Deployment  & Will the predictions be used in decision-making or to trigger an action? How? \\
						& Will the model data be regularly updated? How? \\
						& Will the model be built once, or re-trained continuously?  \\
						& Will the model incorporate feedback? If so, how and what type of feedback? \\
                & {How will uncertainty be communicated? (e.g., are prediction intervals needed?)} \\
				& {What will be done to facilitate} model interpretability/explainability? \\
\hline
\end{tabular}}
    \caption{An overview of modeler decisions when forecasting displacement.}
    \label{fig:model_decisions}
\end{table}
%%%%%%%%%%%%%%%%%%%%%%%%%%%%%%%%%%%%%%%%%%%%%%%%%%%%%%%%%%%%%%%%%%

\section{Prediction problem definition and modeler decisions}\label{sec:decisions}
After having identified and gathered relevant datasets, the next step is to structure the prediction problem itself. We identify nine key decisions that must be made, involving: the unit of analysis, the time horizon, the target variable, the feature variables, the treatment of missing data and data quality, the modeling approach, how models will be selected, how performance will be assessed, how the resulting models will be deployed. These decisions are summarized in Table \ref{fig:model_decisions}.

\subsection{Unit of analysis}
The most basic building block for any model is the unit of analysis: modelers must decide who (or what) will be represented by each item in the dataset.  In their review of the mobility literature, \textcite{toch_analyzing_2019} categorize models according to whether they study users, places, or trajectories. Agent-based models often model refugees and IDPs at the individual (user) level, but in policy settings it is common to work with data at an aggregate (place) level. There appear to be fewer models of displacement at the trajectory level {(i.e., models studying the routes traveled by displaced persons)}, likely because few datasets are detailed enough to support this analysis.

If studying places, modelers must select a level of aggregation. They must first decide whether to work at the national or sub-national level; for example, \textcite{martineau_red_2010} predicts refugee generation by country while \textcite{huynh_forecasting_2020} predict IDP movements by province. If working at the sub-national level, modelers must decide whether to use a country's official administrative regions, or to take a generalizable geospatial approach such as geohashing, which assigns all points on the earth's surface to a hierarchical gridded representation.  Finally, modelers must decide whether to comprehensively cover the entire region of interest, or to focus on specific cities, border checkpoints, and/or refugee camps as in \textcite{suleimenova_generalized_2017}.

In addition to deciding on the geographic coverage of the model, there is the question of which populations will be included in the analysis. For example, it may be necessary to decide whether the model will analyze migrants as well as refugees, although it can be difficult to distinguish between these groups in practice since individuals may cross international borders before being formally granted refugee status. It is also necessary to decide whether to focus on movements that involve crossing national borders, or whether domestic displacement is of interest and IDPs will be included in the sample. Patterns of internal and international displacement may be fairly distinct given restrictions on border crossings; therefore, it may be difficult to create a single model of movement that captures both {types of displacement} well. Ultimately, however, the choice of which populations to include may be determined by the unit of analysis, since e.g. the decision to model flows between countries will necessarily exclude IDPs from the population studied. In other cases, the population studied may be determined by the available data, since information on refugees, IDPs, and migrants may be recorded by separate organizations with different levels of geographic granularity.

\subsection{Time horizon}

Next, modelers must determine the time period over which each unit is observed. If working with a \textit{longitudinal} or \textit{panel} dataset involving repeated observations of the same unit over time, modelers must decide whether units will be measured with daily, weekly, monthly, or annual frequency. Alternatively, the data may arrive in the form of a \textit{cross-sectional} dataset in which each unit is measured only once. One challenge in practice is that different indicators have different levels of temporal granularity; for example, arrivals might be measured on a daily basis whereas food prices might be measured weekly and GDP may only be reported annually. Therefore, it is often necessary to change the resolution of individual datasets, for example by interpolating between measurements or by averaging over several sequential measurements.

Modelers must also decide how far into the future they want to forecast. This decision should be made with the application scenario in mind, with an emphasis on the forecast horizon needed for effective decision making. For example, if humanitarian supplies must be requested six weeks in advance, then it may be advisable to focus on predicting arrivals with a two month horizon. Similarly, if input data for the model arrives with a delay -- e.g., when the input data consists of population surveys that must be entered and processed -- then it may be necessary to predict multiple steps into the future so that data from previous time periods can be used to make predictions even when recent data is not yet available.  In general, predictions farther into the future are likely to be more uncertain, but the amount by which prediction quality degrades with the time horizon may be context-specific and depend on how much the displacement time series ``drifts'' over time.

\subsection{Target variable}\label{sec:target}
%%%%%%%%%%%%%%%%%%%%%%%%%%%%%%%%%%%%%%%%%%%%%%%%%%%%%%%%%%%%%%%%%%
\begin{figure}
\centerline{
\begin{tabular}{|c|c|c|c|}
\hline
Internal Displacement & Total Outflows & Total Inflows & Pairwise flows \\
\includegraphics[height=1in]{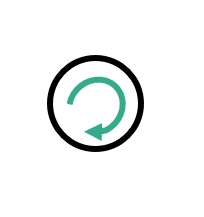} & \includegraphics[height=1in]{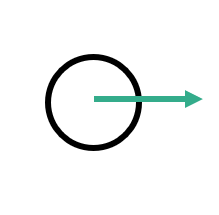} & \includegraphics[height=1in]{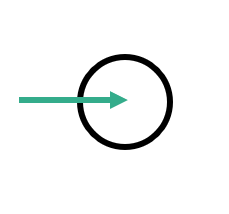} & \includegraphics[height=1in]{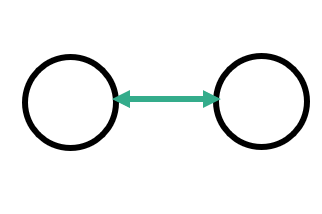} \\
$a_{ii}$ & $\sum_{j: j\neq i} a_{ij}$ & $\sum_{j: j\neq i} a_{ji}$ & $a_{ij}$ \\ 
\hline
\end{tabular}}
\caption{Above, we illustrate four main options for the target variable. We consider a matrix $A$ of pairwise flows for a given time period, where each entry $a_{ij}$ represents the number of people who have moved from region $i$ to region $j$ during that period. Where applicable, the focal region is region $i$.}\label{fig:flow_types}
\end{figure}
%%%%%%%%%%%%%%%%%%%%%%%%%%%%%%%%%%%%%%%%%%%%%%%%%%%%%%%%%%%%%%%%%%

Having specified the unit of analysis and the time horizon of interest, the next decision involves the structure of the target variable. When considering a geographic region of interest $i$, there are four potential quantities of interest (illustrated in Figure \ref{fig:flow_types}): \textit{internal displacement}, or the internal movement within region $i$; \textit{outflows}, or the cumulative departures from $i$ to all other regions $j$; \textit{inflows}, or the cumulative arrivals to $i$ from all other regions $j$; and \textit{pairwise flows}, or the direct movement between $i$ and a given other region $j$. All four quantities can yield useful insights, and the choice between them typically depends on the application setting; for example, humanitarian response teams are typically most interested in arrivals, whereas researchers might be interested in gaining a better understanding of pairwise movement patterns.

Another consideration is whether the target variable should be transformed. For example, \textcite{huynh_forecasting_2020} found that performance was higher when using log counts of IDP flows as the dependent variable. {When models are trained to minimize a given error function, transformations of the dependent variable have implications for how forecast errors are penalized. For example, in absolute terms}, estimating 1,000 instead of 100 yields an error of 900 and estimating 100 instead of 10 yields an error of 90, but in logarithmic terms these errors are equivalent: {$\log_{10}(1,000) - \log_{10}(100) =  \log_{10}(100) - \log_{10}(10)=1$.} Therefore, using log counts rather than absolute counts will place more of the model's ``focus'' on values that are low in absolute magnitude relative to values that are large in absolute magnitude. Displacement time series often see low-level variations interrupted by unexpected bursts or spikes (for an example, see Figure \ref{fig:trends}); in such cases, the use of a log target variable will force the model to focus more on predicting the ``normal'' scenarios, whereas the use of the raw target variable will lead to a greater focus on predicting large {spikes} in the trend.

When pooling data from heterogeneous regions in order to estimate a single model, it is also possible to normalize the target variable by region in order to make the magnitudes more comparable across regions. This may be important in cases where some administrative regions (e.g., those containing key cities) have much larger populations than others; otherwise, models may focus on fitting these high-population regions at the expense of modeling dynamics in smaller regions, since mistakes in modeling high-population regions will lead to prediction errors that are large in magnitude. Another approach to heterogeneity is to define target variables in relative, rather than absolute terms. For example, field teams might be most interested predicting percentage \textit{changes} in displacement counts relative to the previous months. This may address the scaling problem involved in using raw displacement counts while also producing predictions that are of greater operational relevance.

Finally, modelers can consider whether to model displacement as a \textit{classification} rather than a \textit{regression} problem. Regression refers to the attempt to predict a {continuous} outcome, whereas classification refers to the use of a {categorical} outcome. The classes can consist of binary variables (such as whether or not a given region will produce IDPs), or variables with several possible values. For example, it is possible to create a categorical target variable capturing whether IDP counts experience a large increase, a large decrease, or little change; {the definition of ``large'' may depend on the operational context, and may for example represent the number of additional arrivals that would require opening a new shelter}. Such categorical dependent variables are appealing for two reasons: first,  it may be easier to learn a classification model over three categories than a full regression model; and second, producing forecasts in the form of precise integer counts (e.g., ``1,218 arrivals'') may create an undue impression of certainty. For some operational purposes, knowing the general magnitude and direction of changes may suffice for decision making and precise estimates may not be necessary.

While we focus on point forecasts, it is also possible to address uncertainty by predicting a range or expected distribution of the target variable. For example, the forecasts produced by machine learning models can be accompanied by estimated prediction intervals, whereas the predictions of agent-based models can be accompanied by ranges of possible values generated from several different runs of the model, or a sensitivity analysis displaying predictions under different possible assumptions about model parameters. 

\subsection{Feature variables}

After deciding on the quantity to be predicted, the next step is to select the feature variables that will be used for prediction. As described in Section \ref{sec:data}, a growing number of data sources on conflict, politics, economics, the environment, and societal factors are available. These variables can be added to the model according to whether they are believed to correlate with or drive displacement, and whether they are observed with enough frequency and geographic granularity to match the observations on the target variable. Additional features can be manually created, including: binary indicators for region, month of the year, and year; continuous measures of the passage of time; and binary indicators of whether countries share a common border. Context-specific factors that influence movement patterns -- such as indicators for holidays, the school calendar, agricultural planting seasons, or even the schedule of bus trips to the border -- may also be valuable predictors of displacement.

\subsubsection{Incorporating lagged features and features from nearby regions}

Having built a dataset of features by region, modelers can experiment with how these features should be associated with each observation of the target variable. It is possible to incorporate lagged features in order to capture causes that occur further back in time; a displacement driver may only take effect over a long time horizon, in which case previous values of the feature might be informative. For example, environmental factors such as failed rains may have the greatest impact when it comes time to harvest, which could be several months into the future. Lagged values of the target variable can also be included; in particular, 6- and 12-month lags of displacement can be useful if flows are seasonal in nature (e.g., from annual flooding). Unfortunately, \textcite{schmeidl_early_1998} note that ``there is little guidance on how long it takes for certain processes to lead to a humanitarian disaster''; this is an interesting direction for future research.

\begin{figure}
\centerline{
\includegraphics[width=.5\linewidth]{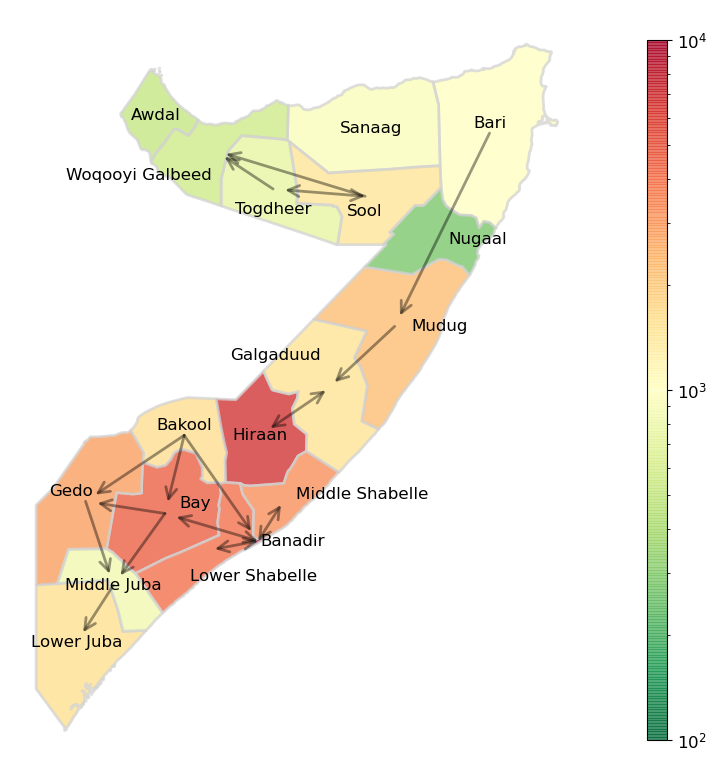}
}
\caption{A heat map indicating the average number of monthly arrivals by region in Somalia during our study period. The arrows indicate pairs of regions which experienced directed flows of at least 100 people per month on average from January 2011 - December 2020, highlighting the most common patterns of displacement. }\label{fig:network}
\end{figure}

Furthermore, there is a question of whether to incorporate feature variables from the focal region(s), from nearby regions, or from all regions. For example, when studying arrivals, observations on the focal region provide information on how attractive that region is; if conflict is low and wages are high in a given location, it might be an ideal destination for displaced persons. However, observations on other regions can provide information on the factors creating the displaced flows, such as fighting or political repression. When deciding how many regions to incorporate, it can be helpful to inspect the extent of displaced flows in the dataset. {In Somalia, we found that many of these flows are local in nature; the most common moves appear to occur between adjacent regions} (see Figure \ref{fig:network}). However, in other settings --- particularly in countries where movement is easier or where the shape of the country is more conducive to traversing movements -- it may be important to use a full set of features from all available regions.

%%%%%%%%%%%%%%%%%%%%%%%%%%%%%%%%%%%%%%%%%%%%%%%%%%%%%%%
%%%%%%%%%%%%%%%%%%%%%%%%%%%%%%%%%%%%%%%%%%%%%%%%%%%%%%%

\subsubsection{Feature selection and model regularization}
Choosing the right mix of features relies on a balance between incorporating expert knowledge about the situation -- i.e., theories about which factors are important for generating and shaping displacement -- and automated processes for reducing or otherwise compressing these features. Although the number of features in displacement datasets tends to be smaller than in many other machine learning applications, reducing the number of features may be desirable for three main reasons. First, if data pre-processing involves the generation of lagged values of different variables or the expansion of the dataset to include features from nearby regions, then the number of feature variables can increase quickly and it may be helpful to identify \textit{which} lags and \textit{which} nearby regions are most informative as predictors. Second, because the size of displacement datasets tends to be small relative to standard machine learning datasets, the number of features can grow large relative to the number of observations and increase the risk of overfitting. Finally, using simpler models and identifying core features may be especially important in displacement applications, where there is often an interest in interpreting models and understanding which features are driving predictions. 

Feature selection and model simplification can be implemented in several different ways. First, it is possible to apply algorithms for automated feature selection (for example, forward or backward feature selection) or to combine the information from many features into a simpler set of features (for example, with Principal Components Analysis). Some machine learning techniques already incorporate built-in methods for feature reduction; for example, random forests automatically select subsets of features during steps in the model construction, whereas deep learning models are able to combine input features into internal \textit{representations} that are themselves highly predictive features. Another common practice in machine learning is to introduce a \textit{regularization} term that penalizes complex models in order to avoid overfitting models to the training data. In general, regularization imposes a cost to selecting coefficients that are large in magnitude, but some types of regularization can also encourage the selection of models with fewer coefficients. Intuitively speaking, flexible models can best fit the data so the incentive when training is almost always to make a model more complex, but a regularization term introduces a cost to this complexity and encourages the use of simpler models. {Finally, modelers may choose to normalize or standardize feature variables to make them more comparable in magnitude. This can improve the performance of some machine learning models.}

\subsection{Missing data, data quality, and data scarcity}\label{sec:missingdata}
{Prior to fitting any} model, it is important to assess data limitations. In their review of the use of artificial intelligence for predicting resource demands in emergencies, \textcite{zhu_comprehensive_2019} {identify data limitations including scarcity, inaccuracy, and incompleteness}. In our experience, displacement data suffers from all of these problems. {With regard to scarcity, }while the availability of data is growing, some of the primary challenges include the coverage {and interoperability of the data. On the other hand, uncertainty can arise from inaccuracy or quality issues in the measurement of the target or feature variables, and even from attempts to handle incomplete data (for example, by imputation) as described below. Uncertainty can be further amplified when models are built on this flawed data. }

 Because many displacement-related datasets are collected by humanitarian agencies, they are often limited to certain populations -- such as refugees \textit{or} IDPs -- which are of interest for operational purposes. Data collection tends to be incident- and crisis-specific, and there is a need for more systematic data collection across multiple emerging scenarios. Finally, most datasets focus on a specific class of indicators -- for example, economic or political indicators -- and different datasets must therefore be integrated to produce a modeling dataset. This aggregation can be the most costly and time-consuming part of the modeling process.

Another important consideration in the context of forced displacement is the presence of missing data. In many cases, scarce or missing data is correlated with the outcome of interest; for example, we may lack observations on precisely those regions that suffer the most displacement, because it is hardest to collect data in those areas. This effectively shifts \textit{what} we are able to predict: instead of capturing all displacement across all regions, we are capturing \textit{measurable} displacement across regions \textit{to which enumerators have access}. This prediction problem is also relevant; if enumerators cannot access a conflict region, it will be challenging for humanitarian aid to reach that region even if displacement is occurring. However, this is an important reframing of the problem that should be understood and acknowledged by all system users, as it changes the focus of prediction efforts. 

To address missing values, different approaches are available to modelers. Where values of the target variable are missing, it may make sense to drop missing values, although this may bias the dataset as described above. Where values of the feature variables are missing, a common alternative strategy is to impute missing values -- for example, by carrying past values forward or using historical means -- although it is important to avoid ``data leakage'' by ensuring that imputation is done using \textit{only} data that will be available at the time of prediction (i.e., only data from the current or preceding time stamps). A final alternative involves coding binary indicators to flag whether an observation is missing or not. The latter approach may be particularly appealing in displacement contexts, since if this ``missingness'' is not random (e.g., for the reasons described above) it might actually be informative for the model.

\section{Model implementation}\label{sec:implementation}

\subsection{Modeling approach}\label{sec:models}
Once data has been collected, aggregated, and structured for the learning problem, the next step is to select the method used to forecast displacement. Options include econometric {and statistical forecasting} models such as multiple regression, gravity, and time series forecasting models; agent-based models; and machine learning and deep learning models.

\subsubsection{Econometric and statistical forecasting models}
Some of the earliest models of displacement use multiple regression models to estimate factors that predict displacement.  For example, \textcite{martineau_red_2010} fits the following logistic regression model:
\begin{eqnarray}
\log \left( \frac{\Pr(\text{refugees}=1)}{\Pr(\text{refugees}=0)}  \right) = \alpha + \beta X 
\end{eqnarray}
where the dependent variable is the log odds that a country generated over 5,000 refugees between 1996 and 2005, $\alpha$ is a constant, and the set of predictors $X$ includes different political, economic, and demographic factors. The coefficients of interest are the $\beta$, which capture the estimated effect that these input variables have on the log odds of producing refugees. 

Such models are appealing because they are simple and interpretable, and because they facilitate exploration of key factors driving displacement. However, these models often lack the sophistication to model more complex relationships between variables, or the expressiveness to capture spatial and temporal patterns. Gravity models extend these simple regression models to capture spatial flows, whereas time series forecasting models extend these simple regression models to {better model how series evolve over time.} 

\textit{Gravity models} are designed to work with a network/graph representation of flows, and focus on predicting the volume of flows between an origin and a destination. They are a popular tool in the migration and international trade literature, in which case flows might consist of e.g. the number of people or the volume of goods moving from point $i$ to point $j$. Gravity models assume that the flow between two points is a function of the features of both points, as well as the distance between them:
\begin{eqnarray}
a_{ij} = \alpha \frac{X_i ^{\beta_1}\ X_j^{\beta_2}}{d_{ij}^{\beta_3}} 
\end{eqnarray}
which is often estimated in practice using the transformed equation:
\begin{eqnarray}
\log(a_{ij}) = \alpha + \beta_1 \log(X_i) + \beta_2 \log(X_j) - \beta_3 \log(d_{ij}) 
\end{eqnarray}
where $a_{ij}$ indicates the size of the flow between $i$ and $j$; $X_i$ includes characteristics of the origin and $X_j$ includes characteristics of the destination; and $d_{ij}$ represents the distance between the two points. One of the benefits of these models is that they explicitly model features of the origin and destination that act as push and pull factors, respectively; they also allow modelers to impose a cost for the trip between two points. A key challenge of this approach, however, is the need to have comprehensive data on pairwise flows, which is sometimes hard to acquire relative to counts of arrivals or departures.

Another alternative to simple regression models consists of time series forecasting models. These models build on four key ideas: the use of \textit{autoregressive} terms, i.e. lagged values of the target variable, as predictors; the use of \textit{moving averages} to capture long-term trends in the target variable; the use of \textit{differencing} to capture short-term changes in the target variable; and the modeling of \textit{seasonality}, which might be used to account for repeated patterns in the time series. These models are popular in forecasting applications such as financial modeling. However, somewhat surprisingly, we have not identified any papers applying these models to forced displacement.

\subsubsection{Agent-based models}

ABMs represent another class of models that has recently gained popularity in modeling a variety of refugee and IDP scenarios \cite{ suleimenova_generalized_2017, sokolowski_modeling_2014, crooks_agent-based_2014,anderson_simulation_2007,aylett-bullock_operational_2021}. These models specify an environment in which agents take actions at sequential timesteps. Depending on how an agent's actions are motivated or constrained, different patterns of behavior emerge naturally; these patterns can be explored through the use of simulations. In biological applications, for example, modelers have been able to replicate the ``flocking'' or ``swarming'' behavior commonly observed in animal groups; one of the appealing aspects of ABMs is that complex behaviors can emerge from fairly simple environments and behavioral rules. An early application of ABMs is the Schelling Model, which describes how patterns of residential segregation can emerge even when individuals do not explicitly prefer to live in areas in which a majority of residents share their own ethnic affiliation \cite{schelling_dynamic_1971,zhang_tipping_2011}; this model has been applied in a migration context to analyze how communities respond to the influx of migrant groups \cite{urselmans_schelling_2016}.

ABMs are appealing in this context because they can propose a behavioral model of displacement processes, and allow for the simulation of outcomes under different conditions and assumptions. However, one challenge with ABM approaches is that it is difficult to validate these models. Authors typically assert that their model works if the reproduced behavior approximately matches observed data, but these models are generally designed to capture a theory of behavior rather than to optimize predictive performance. Multiple competing models of behavior may produce similar predictions, and just because a model is currently calibrated to reproduce past observations does not mean that it will successfully predict future observations. ABMs may therefore show more promise for developing insights into system dynamics, relative to making accurate predictions.

\subsubsection{Machine learning and deep learning models}\label{sec:ml_dl}
Another approach to prediction involves the use of machine learning models. Typically, these models are categorized along two dimensions: (1) whether they involve classification or regression as described in Section \ref{sec:target}; and (2) whether they are supervised or unsupervised learning methods. A \textit{supervised} learning problem is one for which there are known ``answers'', or labels; these labels are used to train the algorithm by penalizing any predictions that deviate from the ground truth. This setting matches most displacement forecasting problems, since information about past flows can typically be used to train the model for the future. In contrast, \textit{unsupervised} learning identifies patterns or groupings in the data without knowing the ``true'' labeling; an application to displacement forecasting could involve identifying clusters of regions that see similar displacement trends over time.

There are a number of popular supervised machine learning algorithms that can be used for regression or classification problems. For example, \textit{lasso} and \textit{ridge} \textit{regressions}  are linear regression models that incorporate a {regularization penalty} on the coefficient weights of the model; this allows the regression to consider a large number of variables while constraining overfitting. \textit{Support vector machines} refer to a classification approach that attempts to separate observations using a hyperplane between them; the algorithm penalizes the observations that fall on the wrong side of the divide, and the penalty increases with the distance from the dividing hyperplane. \textit{Random forests} \cite{breiman_random_2001} are models that are built from a set of many decision trees; in {an attempt} to prevent the trees from overfitting, the model selects a new sample of observations for each tree, and a new subset of variables to form each split in the tree. The predictions of individual trees are then averaged together in an \textit{ensemble}. Yet another group of models includes boosting algorithms such as \textit{AdaBoost} \cite{freund_decision-theoretic_1997} and \textit{XGBoost} \cite{chen_xgboost_2016}; these are also ensemble models, but the component models are built sequentially such that each tries to correct the errors of the previous one. An accessible introduction to many of these methods can be found in \textcite{james_introduction_2013}.

In practice, it is common to test many different models because there is not a strong prior regarding which will work for the problem at hand (for example, this is the approach adopted by \textcite{huynh_forecasting_2020} in their model of internal displacement, and \textcite{nair_machine_2020} in their model of mixed migration).  Often, sophisticated models such as random forests, Adaboost, and XGBoost are the top-performing models because they match the data well due to their flexibility. However, one concern with all of the machine learning methods described above is that they do not explicitly incorporate a historical component, unless lagged features are manually engineered and added to the model. 
  
To better model sequences, many researchers have begun to rely on deep learning methods, which have made advances in sequence modeling due to the field's interest in audio, video, and text data. Two popular types of sequence models are Recurrent Neural Networks (RNN) and a variant of these networks called Long Short-Term Memory (LSTM) networks {\cite{hochreiter_long_1997-1}}. RNNs allow for dependence between subsequent observations, {enabling} previous units' values to influence the current unit's predictions. However, one challenge with RNN approaches is that as an observation is farther and farther back in time, it becomes less likely that it will influence the current prediction. LSTMs overcome this by adding a capacity for ``memory'', allowing historical values to propagate forward more easily.

One appealing aspect of these networks is their adaptable structure; for example, multivariate LSTMs can be used to simultaneously predict multiple time series, or multi-step LSTMs can be used to forecast multiple time steps into the future. However, challenges involved in using neural networks include the computational cost of training these networks, which may make them difficult to deploy in resource-constrained settings, as well as their high data requirements, which may make them unsuitable for shorter {displacement} time series. In particular, these models can easily overfit when trained on small samples (a problem we observe in our own setting, as demonstrated in Table \ref{fig:ml_results}). {In general,  researchers using deep learning methodologies have} attempted to address these problems through the use of \textit{data augmentation} techniques to generate synthetic samples, as well as \textit{transfer learning} to adapt models learned in one setting to a new context. While the concept of data augmentation seems promising given the limited datasets available on displacement, the success of transfer learning will depend on finding analogous problems that have well-developed models built on more extensive datasets.

Conceptually speaking, it is important to note that the primary focus of machine learning to date has been \textit{prediction} rather than \textit{causal explanation}.\footnote{For further reading on this topic see \cite{breiman_statistical_2001, shmueli_explain_2010, mullainathan_machine_2017}. In recent years there has been a dramatic growth in research on machine learning approaches for causal inference, so this distinction is not as clear as it once was.} This is in contrast to econometric and agent-based models, {which are generally constructed according to} theories about the structural mechanisms and decision processes underlying displacement. In machine learning, model construction tends to proceed backwards from the end goal -- the quantity to be predicted, or the {target} -- and models are chosen in order to minimize a given {error metric} that penalizes differences between predictions and the true target value. Once a prediction target and error function are chosen, researchers are fairly agnostic to the choice of predictive {features} as well as to the functional form of the model itself.

One of the resulting {characteristics} of machine learning models is that they tend to offer principled and pragmatic approaches to automated model and feature selection, which means that they may rely less on theoretical assumptions about how displacement processes occur and which features are important. However, a common criticism of machine learning models is that they are ``black box'' models which lack interpretability; it may be necessary to conduct additional analysis{, or select families of explainable machine learning algorithms in the design stage, in order} to identify important features and how they influence the {resulting forecasts}.

\subsection{Model selection}
The next step in the modeling pipeline is to develop a strategy for selecting optimal parameters for the proposed models. In general, machine learning models can capture very sophisticated relationships between feature variables and the target variable, so these models may easily overfit to a given dataset. For this reason, researchers typically split their datasets into multiple parts and use these parts to train and evaluate their models. 

When modelers have access to large amounts of data, they may split the data into three parts: a \textit{training} dataset used to fit some candidate models; a \textit{validation} dataset used to choose among these models; and a \textit{test} dataset to assess the chosen model's performance on completely new data. When less data is available, as may be common in forced displacement contexts, researchers often perform training and validation on a single dataset using a \textit{k-fold cross-validation} strategy. For example, a 10-fold cross-validation strategy would partition the dataset into 10 different subsets. Then, each candidate model would be fit 10 times; in each iteration, 9 of the subsets would be used for training and the 10th would be used for assessing performance.  When working with time series data such as arrival or departure trends, cross-validation strategies {may} make accommodations for the sequential nature of the data, so that data from the future is not used to build models which predict the past. Figure \ref{fig:tscv} shows two common approaches to cross-validation in this context, using sliding and expanding cross-validation windows. Relative to standard cross-validation, time series cross-validation is constrained: either a small amount of data is used to fit each fold (Figure \ref{fig:tscv_c}), or earlier data is over-represented in the folds (Figure \ref{fig:tscv_b}).

\subsection{Model performance}\label{sec:performance}
Once an approach to model selection has been developed, the next step is to consider strategies for assessing model performance. Typically, this involves two choices: deciding on performance and/or evaluation metrics for scoring the models, and specifying clear and reliable benchmarks of what baseline performance should look like.

\subsubsection{Error metrics}
In order to train prediction models and compare different models, it is necessary to identify an approach to quantifying model performance. In practice, there are several popular error metrics for regression models, including mean squared error (MSE), mean absolute error (MAE), and mean absolute percentage error (MAPE); each of these scoring methods shapes model choice in different ways. For example, using MSE rather than MAE means that large prediction errors will be penalized with extra weight (due to the use of the squared term, which increases the weight given to large values). Selecting MAPE as the scoring methodology may give more weight to regions with small numbers of arrivals, since e.g. predicting 150 arrivals instead of the true value of 100 will be penalized just as heavily as predicting 15,000 arrivals instead of the true value of 10,000. The question of which of these errors should be penalized more heavily will likely depend on the operational context envisioned by the modeler. For example, it is clearly more difficult to prepare supplies for 5,000 new arrivals than it is for 50 new arrivals, so it may be desirable to focus on correctly predicting large values. However, a region that typically receives around 10,000 arrivals might already be prepared for new influxes of IDPs, whereas a region that typically receives a lower number of arrivals might not be prepared to handle even a modest 50-person increase, in which case it may be preferable to focus on predicting unexpected increases in low-arrivals regions.

One challenge in selecting the appropriate error metric is capturing the ``burstiness'' and spikes in many displacement time series; for example, the number of people displaced may escalate quickly in the event of natural disasters or conflict outbreaks. Since different error metrics penalize extreme values in different ways, the choice of metric will influence the tendency of models to capture anomalies in the data. However, it may also be valuable to explicitly compare how well different models capture these periods of interest, for example by assessing model performance on a restricted set of the data that only includes these unusual periods. An interesting area for future research is whether models for extreme events -- which have been developed in fields such as environmental and financial modeling -- may be adapted to forced displacement settings.

Relatedly, we note that practical and political considerations may favor the use of asymmetric loss functions which penalize overestimates more or less aggressively than underestimates. For example, in some cases over-prediction may be worse than under-prediction: if arrivals are overestimated, then humanitarian organizations may incur a financial expense to move resources unnecessarily or divert resources from existing emergencies, whereas under-prediction carries less risk because it does not trigger any concrete action. On the other hand, the failure to anticipate crises and respond may act against humanitarian principles which mandate that ``human suffering must be addressed wherever it is found'' \cite{ocha_what_2012}, and under-prediction is costly from this perspective; \textcite{altay_forecasting_2020} give an example of a practitioner who would rather destroy {an excess of} unused, expired medical supplies than risk a shortage {by underestimating the need for supplies}. \textcite{van_der_laan_demand_2016} note that MAPE is already an asymmetric loss function that penalizes over-predicting because under-predictions can have a maximum error of 100\%, whereas over-predictions can accrue very large errors. For other standard loss metrics such as MSE or MAE, a simple approach to implementing asymmetric loss functions is to add an additional multiplier that scales the loss of over-predictions relative to under-predictions.

In the end, the choice between error metrics often amounts to a decision about whether to prioritize prediction quality for periods when displacement flows are low vs. high in absolute value. However, one can imagine that certain models systematically perform better across all regions, regardless of the average amount of displacement in each region. In practice, we found that it can be difficult to identify such models, since when examining errors across models and regions the difference in magnitudes \textit{between} regions often dwarfs the differences in magnitudes across models \textit{within} regions. One strategy that we have applied is to rank model performance by region. In this way, we can manually inspect the data and determine whether certain models are consistent ``winners'' across different geographic areas (see Figure \ref{fig:rank}). Of course, an alternative approach could simply involve developing different models for each region. A challenge with this approach is that the amount of historical data per region is often limited, and so the data available at the regional level may not suffice to train high-performing models. Therefore, we feel that it is most promising to invest in strategies for training models on pooled datasets as described here.

\subsubsection{Model benchmarks}
Once an approach to scoring models has been selected, modelers should assess whether their forecasts perform better than an alternative or default approach to prediction. A popular choice of benchmark consists of ``na\"ive'' predictions. For example, \textcite{huynh_forecasting_2020} use the last observation carried forward (LOCF) and historical means (HM) as their baseline models. \textcite{suleimenova_generalized_2017} also consider linear extrapolation from two data points -- the starting and the current period -- as well as ``extrapolation by ratio'', which refers to the assumption that the distribution of refugees over destinations will remain constant even as the number of refugees increases. One challenge is that there are many different possible baselines to consider (for example, we can carry observations forward with different lags, and calculate different types of means including expanding means, exponentially weighted means, and historical means with different windows) and so even the optimal baseline model is something that can be ``learned'' from the data. 

While heuristic benchmarks offer good reference points, a more fundamental challenge to evaluation is that the \textit{true} counterfactual benchmark -- that is, the prediction that would be made by teams on the ground in the absence of any predictive analytics tool -- is often an expert guess. It is relatively easy to identify failures of any given predictive model, but to assess the usefulness of the model, these failures should be compared to a humanitarian organization's default decisions and recommendations.  This can be challenging because many organizations do not formally document their decision-making processes; however, it is worth attempting to understand how expert guesses and the corresponding resource allocation decisions are actually made; which factors (written and unwritten) and criteria are taken into consideration; and even which forecasts or decisions have actually been made in the past. When this information can be obtained, it may provide the best means for assessing the value added by predictive models.

\subsection{Deployment}
A final consideration involves how predictive models should be deployed. In the context of humanitarian operations, deployment involves logistical considerations such as who will own and maintain the models developed, and whether model development and dissemination can occur in resource-constrained settings that may have limited computational power or access to the internet, or whether these models must be centrally managed. It is also necessary to plan for how models will be adapted based on new information.  This includes whether models will be re-trained as time passes and new data arrives; how to flag any declines in performance that might occur because models no longer represent the situation on the ground; and how often the model data and predictions will be updated. It may be advisable to develop a pipeline for the collection and ingestion of new data at regular intervals.

Furthermore, there is the question of how predictions should be communicated and displayed to practitioners.  
In order for deployment to be successful, there is the need to build trust in the forecasts. In part, this means ensuring that models are not deployed until performance is sufficient, and that practitioners are given tools for better understanding the models \cite{andres_scenario-based_2020}. These can include opportunities to interrogate the model through the use of interactive data visualizations of the predictions and results, as well as the use of model explainability strategies such as variable importance weights and SHAP values \cite{lundberg_unified_2017}. 
Building trust also requires that uncertainty around a model and its predictions is communicated carefully. Strategies for communicating uncertainty include: providing prediction intervals around a forecast (and/or confidence intervals around parameter estimates); displaying predictions in terms of a range rather than a single value; and displaying predictions from several top models rather than the single best performer. 

%%%%%%%%%%%%%%%%%%%%%%%%%%%%%%%%%%%%%%%%%%%%%%%%%%%%%%%%%%%%%%%%%%
\begin{figure}
\centerline{
\includegraphics[width=\linewidth]{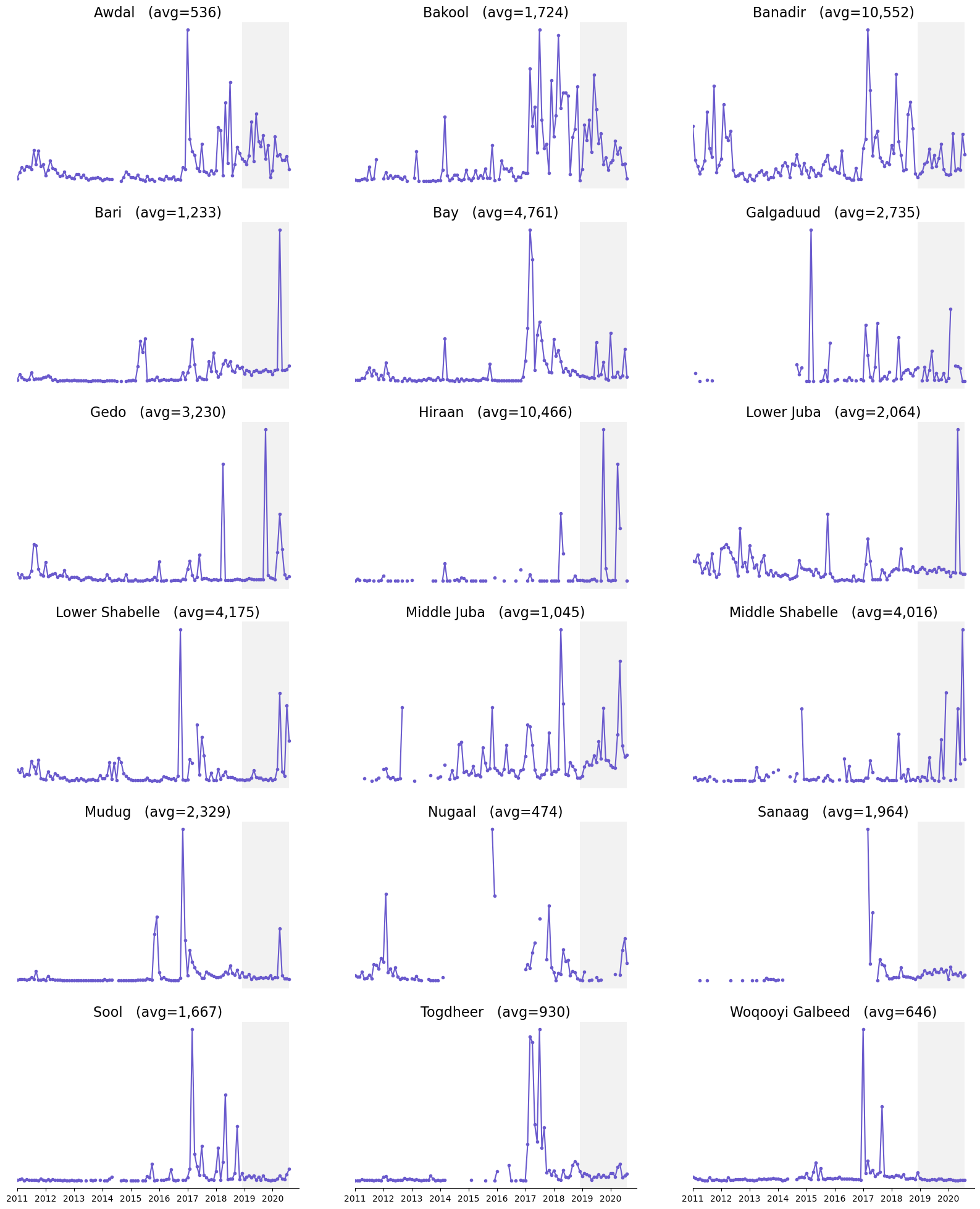}
}
\caption{Arrivals by region in Somalia. The series have been rescaled such that the trends are comparable in magnitude, but note that there are wide disparities in the average numbers of arrivals per region, and that large periods of missing observations are evident for some regions. 
The training {period} used for machine learning is illustrated with a white background, whereas the test {period} is shown in grey.
}\label{fig:trends}
\end{figure}
%%%%%%%%%%%%%%%%%%%%%%%%%%%%%%%%%%%%%%%%%%%%%%%%%%%%%%%%%%%%%%%%%%

\section{Case study: Somalia}\label{sec:somalia}
We now provide a more detailed discussion of our experience developing displacement forecasting models in Somalia as a means of illustrating the framework above. This forecasting was done in support of Project Jetson, an initiative of UNHCR Innovation which was designed to pioneer the use of predictive analytics for humanitarian operations using Somalia as a case study~\cite{earney_pioneering_2019, unhcr_innovation_predictive_2019}. 
Somalia remains a priority country for predicting humanitarian needs due to the severity of displacement driven by a combination of geopolitical, economic, and environmental factors. As of December 2018, out of a population of 12.3 million people, an estimated one-fifth (2.6 million) were displaced and one-third (4.2 million) were in need of humanitarian assistance. The primary causes of displacement are conflict (33\%), drought (22\%), economic factors (16\%), and evictions (5\%) \cite{unocha_2019_2018}. 

\subsection{Unit of analysis and time horizon}
After consultation with its humanitarian staff in Somalia and an assessment of the available data, the UNHCR Innovation team determined that the project should focus on making monthly predictions for each of Somalia's 18 regions with a 1-3 month forecast horizon. This choice was made in part due to data constraints (many datasets were available only on a monthly basis) and in part due to practical concerns (a 1-3 month horizon was considered the appropriate lead time for preparing a response). UNHCR identified approximately ten years of historical displacement data for use in training the model, making Somalia a promising setting for a predictive analytics project; we note that such a large amount of historical data is rarely available in displacement contexts.\footnote{ Iraq (\url{https://dtm.iom.int/iraq}) and Mali (\url{https://dtm.iom.int/mali}) are two other promising contexts in which IOM has been tracking displacement since 2014 and 2013, respectively.}

\subsection{Target variable}
The target variable was the number of arrivals by region and month, which included both inflows from other regions and internal displacement within the focal region. This choice of target variable was made because the motivation for creating the predictive model was to better serve the needs of all newly arriving IDPs in a region. For context, Table \ref{fig:flows_and_missing} shows average internal displacement, outflows, and inflows by region in Somalia; interestingly, we note that almost all regions experience a nontrivial amount of internal movement.

Arrivals were calculated from UNHCR's Protection and Return Monitoring Network (PRMN) dataset \cite{unhcr_somalia_2019}. The PRMN is a network of enumerators and observers from various partner agencies on the ground in Somalia. Using field interviews of arrivals at checkpoints and IDP camps, PRMN gathers data on: the number of arrivals; their home, previous, and future districts; their reason for displacement; their humanitarian needs; and their demographic composition by gender and age group. {The PRMN dataset was used for analysis because it represents the most comprehensive sub-national data available for prediction; however, two considerations are worth noting}, as they may be common to many datasets. First, data on IDPs can be highly sensitive; PRMN data were initially shared on a monthly basis in a secure file, and had to be aggregated before they could be moved to a shared server. This imposes a constraint on the reproducibility and potential openness of the full data analysis pipeline. Second, the PRMN dataset is not comprehensive; data collection depends on the location of field observers, and gaps in the data can occur when enumerators cannot gain safe access to conflict areas \cite{unhcr_prmn_2017}.

Time trends and a heatmap of monthly arrival counts are shown in Figures \ref{fig:trends} and \ref{fig:arrivals}, respectively. On average, we observed 3,036 arrivals per region and month, with a median of 657 arrivals and a standard deviation of 10,151. As mentioned above, the distribution of arrivals is highly skewed; for example, the average monthly arrivals of displaced persons varied from 474 in Nugaal to 10,552 in Banadir (which contains Mogadishu, {the country's capital}). 

\subsection{Feature variables}
The dataset of features used for prediction included conflict incidents and fatalities for violent events and demonstrations from the ACLED API; data on average commodity prices by market and month from FSNAU; data on rainfall, river levels, vegetation cover, and illness from the FSNAU Early Warning/Early Action dashboard; and estimated driving distances between each region pair from OSM (for more discussion of these data sources, see Section \ref{sec:data}). We also manually engineered a number of additional features; a full list of the variables used in the model shown in Table \ref{fig:data_sources}. By plotting the relationships between these feature variables in Figure \ref{fig:corrplot}, we can see that variables of a given type tend to be correlated: for example, water levels across all rivers tend to move together, as do the economic variables.  In general, though, features were selected to ensure that each included variable contributes distinct information. {The feature variables were normalized (in the case of standard machine learning models) or standardized (in the case of the LSTM) prior to fitting the models.}

\subsection{Missing data, data quality, and data scarcity}
Because each observation in the disaggregated PRMN dataset consists of a single arrivals record, it is difficult to distinguish between periods with no arrivals (zero records were created) and missing arrivals (no observer was recording, due to reasons of security or a lack of enumerator capacity).  Since low monthly arrivals counts were rare in our dataset, we treated all region-month observations with zero recorded arrivals as missing values.

This resulted in a lot of missing data for the target variable: the worst region, Sanaag, was missing data for {53\%} of observations of the target variable (see Table \ref{fig:flows_and_missing}). Missing values of the target variable were dropped in training, so that models only learned from complete records; otherwise, we would have faced the risk of training models that were simply learning how missing values of the target variable were imputed. 
One shortcoming of this approach is that it may shift the modeling focus away from observations of interest, since observations with missing data may represent precisely those regions and periods that experience high insecurity and therefore have high volumes of displacement. 
However, it is not obvious how to overcome this limitation without collecting better data. 

For feature variables with missing values, binary indicators were created to flag whether a variable was missing or not. Then, missing values were replaced by propagating the last recorded value forward, so that the imputation of missing values would use only data available at the time of prediction. All remaining missing values (for which there was no preceding historical data) were filled with a value of zero.

\subsection{{Modeling approach, model selection, and model performance}}

We pooled data from all 18 regions and time periods to fit a single displacement model. This stands in contrast to earlier experiments, in which a separate model was fit for each region.The decision to use an aggregate model was made in order to (1) reduce the risk of overfitting to individual regions; (2) expand the dataset available for modeling (from $n=116$ observations per region to $n=116\times 18$ observations across all regions); and (3) to reduce the administrative burden of fitting and updating models for 18 separate tasks.

The final input dataset covered the months from January 2011 to August 2020, yielding a total of 116 months of observations across 18 regions. We trained our models using data for the forecast horizon extending through December 2018, while the test data covers the forecast horizon from January 2019 to August 2020. For parameter fitting, we implemented 10-fold time series cross-validation with an expanding window approach. 

Our primary models included a range of standard regression algorithms, namely: ridge and lasso regressions, multi-layer perceptrons, XGBoost and AdaBoost, decision trees, random forests, and a Long Short-Term Memory (LSTM) neural network. Although we initially experimented with proprietary tools -- such as Eureqa and H20.ai -- we ultimately decided to proceed with open-source models implemented in Python for reproducibility and intelligibility. We benchmarked our performance against four na\"ive baseline models:
\begin{itemize}
    \item \textbf{Lagged values} which simply assume that arrivals will be equal to the observed arrivals value from $n$ periods ago; for a lag length of 1, this approach is also referred to as last observation carried forward (LOCF)
    \item \textbf{An expanding mean (EM)}, which is an average over all data observed to date  (i.e. not restricted to recent months)
    \item \textbf{An exponentially weighted mean (EWM)}, which is an $n$-month ``average'' which gives more weight to recent periods
    \item \textbf{A historical mean (HM)}, which is simply a rolling average of the last $n$ months' arrivals
\end{itemize}
In the above cases, $n$ was selected according to performance on the training dataset; we chose 1- and 12-month lags, {8 and 23}-month horizons for the exponentially weighted mean, and a {12}-month horizon for the historical mean. 
%%%%%%%%%%%%%%%%%%%%%%%%%%%%%%%%%%%%%%%%%%%%%%%%%%%%%%%%%%%%%%%%%%

\begin{table}
\centerline{
\begin{tabular}{lllll}
\toprule
{} & \multicolumn{2}{l}{1-month horizon} & \multicolumn{2}{l}{3-month horizon} \\
            &           Train &  Test &           Train &  Test \\
\midrule
Perceptron         &            6660 &  6288 &            6811 &  6669 \\
Ridge Regression   &            6152 &  6389 &            6709 &  7015 \\
Expand. Mean       &            7683 &  6437 &            7767 &  6512 \\
Lasso Regression   &            6084 &  6449 &            6457 &  7267 \\
Random Forest      &            5760 &  6471 &            5564 &  6754 \\
Exp. Wt. Mean (23) &            7303 &  6546 &            7600 &  6674 \\
LSTM               &            1377 &  6555 &            1449 &  7195 \\
Exp. Wt. Mean (8)  &            7195 &  6601 &            7794 &  6702 \\
AdaBoost           &            5424 &  6631 &            4962 &  7361 \\
Hist. Mean (12)    &            7428 &  6632 &            7739 &  6774 \\
XGBoost            &            7239 &  6965 &            5849 &  6749 \\
Decision Tree      &            6791 &  7853 &            6421 &  6936 \\
12-month lag       &            9167 &  8487 &            9150 &  8537 \\
1-month lag        &            8294 &  8594 &               - &     - \\
\bottomrule
\end{tabular}
}~\\
\caption{~Root Mean Square Error (RMSE) across different models tested. From the errors, we can see that more flexible models such as random forests and Adaboost are overfitting, and that this problem is particularly pronounced for the sequence model tested (an LSTM). In order to ensure comparability across models, average RMSE was calculated only over observations for which all models produced a prediction. For this subset of observations, the mean number of arrivals was 3,255, with a standard deviation of 7,700 and a maximum of 76,267.}\label{fig:ml_results}
\end{table}
%%%%%%%%%%%%%%%%%%%%%%%%%%%%%%%%%%%%%%%%%%%%%%%%%%%%%%%%%%%%%%%%%%

Results are shown in Table \ref{fig:ml_results}. The best parameters for each type of {machine learning} model were selected through cross-validation on the training dataset, and the performance of these models over the pooled training dataset is shown in the left-hand columns of the table. These models were then applied to produce predictions on the unseen test dataset, which are scored in the right-hand columns of the table. We note that average RMSE was calculated only over observations for which all models produced a prediction; this was done to ensure the comparability of RMSE across models.\footnote{Since the LSTM used a 12-month window of historical arrivals to produce predictions, it did not produce forecasts for 2011 and therefore no observations from 2011 were included in the calculation of the RMSE. Furthermore, observations were dropped when a lagged value of the target variable was missing for one of the prior 23 months; as a consequence, model scoring focuses on periods and regions with relatively stable data collection.}
As we can observe, the RMSE from the models was generally around 6,000-7,000, which is large relative to the average number of 3,255 arrivals for the associated observations, but small relative to the maximum number of arrivals for the associated observations (76,267). 
  
In general, we found that na\"ive benchmarks (particularly, the expanding mean) were competitive with machine learning results; the perceptron and ridge regression were able to beat this benchmark for a 1-month prediction horizon, but the expanding mean was the top-performing model for the 3-month horizon. At the same time, the simplest benchmarks (the 1-month and 12-month lags) performed much worse, suggesting that experimenting with the choice of benchmark is valuable if we truly want to quantify the value added by machine learning models. Finally, we note that some of the more flexible machine learning models -- such as random forests and AdaBoost -- appear to have suffered from overfitting. This was particularly clear in the case of the LSTM, which achieved a very good fit on the training data but performed much worse on the test dataset.

%%%%%%%%%%%%%%%%%%%%%%%%%%%%%%%%%%%%%%%%%%%%%%%%%%%%%%%%%%%%%%%%%%
\begin{figure}
\centerline{
\includegraphics[width=0.6\linewidth]{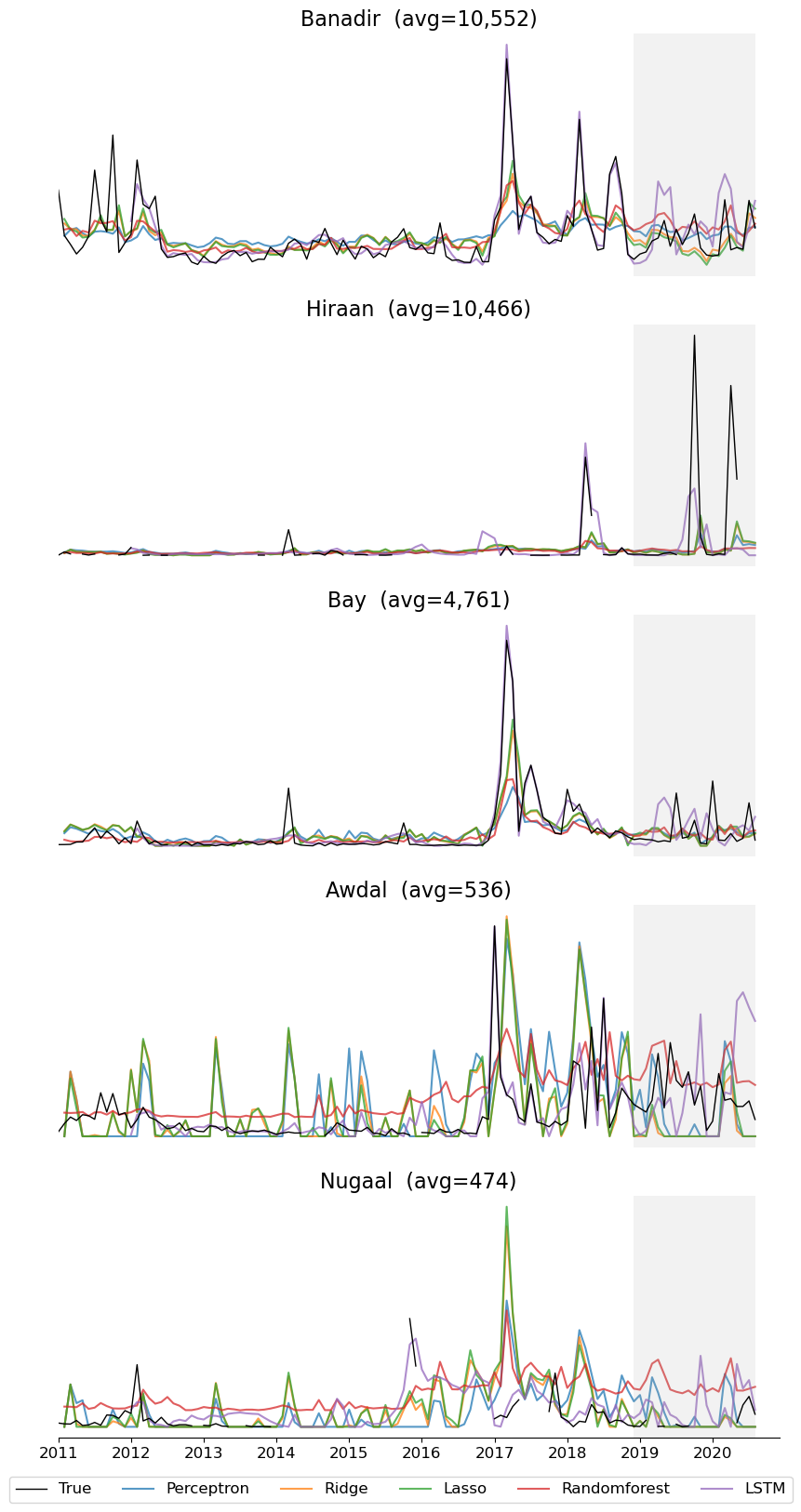}
}
\caption{Illustration of model predictions for the regions with the highest average number of arrivals (Banadir, Hiraan, and Bay) and the lowest number of arrivals (Awdal and Nugaal). Arrivals by region in Somalia are shown in black, indicating the ``ground truth.'' The series have been rescaled such that the trends are comparable in magnitude. The top-performing predictive models on the test dataset for the one-month forecast horizon are shown in color. The training {period} is illustrated with a white background, whereas the test {period} is shown in grey.
}\label{fig:preds}
\end{figure}
%%%%%%%%%%%%%%%%%%%%%%%%%%%%%%%%%%%%%%%%%%%%%%%%%%%%%%%%%%%%%%%%%%

To further illustrate the machine learning models, sample predictions made by the five best-performing models (the multilayer perceptron, the ridge and lasso regressions, the random forest, and the LSTM) on the regions with the highest and lowest numbers of arrivals are shown in Figure \ref{fig:preds}. While the models appear to follow the general trends in arrivals, we can see that they generally do not capture the full extent of bursts or spikes in arrivals, even in the training dataset. The LSTM is able to better capture these unusual periods, but this appears to be because it has overfit to the data.

As described in Sections \ref{sec:target} and \ref{sec:performance}, one challenge with fitting models in this setting is that the level of arrivals is highly disparate across different regions. Therefore, a metric like MSE will penalize small relative errors in high-volume regions much more aggressively than similar errors in low-volume regions, and could potentially lead to the choice of algorithms that perform poorly in low-volume regions. This can be seen in Figure \ref{fig:preds} in the case of Awdal and Nugaal, which have low average numbers of arrivals and very noisy predictions. 

\begin{figure}
\centering
\begin{subfigure}{\textwidth}
\centerline{\includegraphics[width=.8\linewidth]{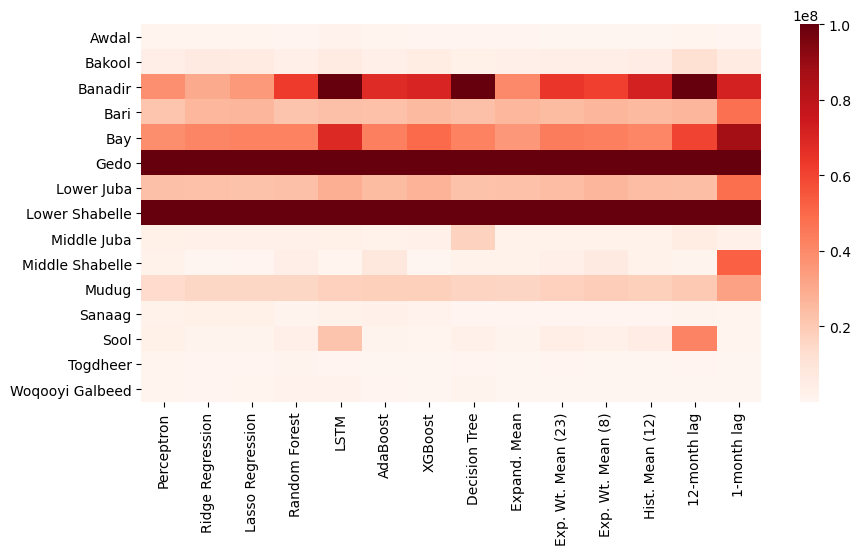}}
\caption{Raw MSE}
\end{subfigure}
\begin{subfigure}{\textwidth}
\centerline{\includegraphics[width=.8\linewidth]{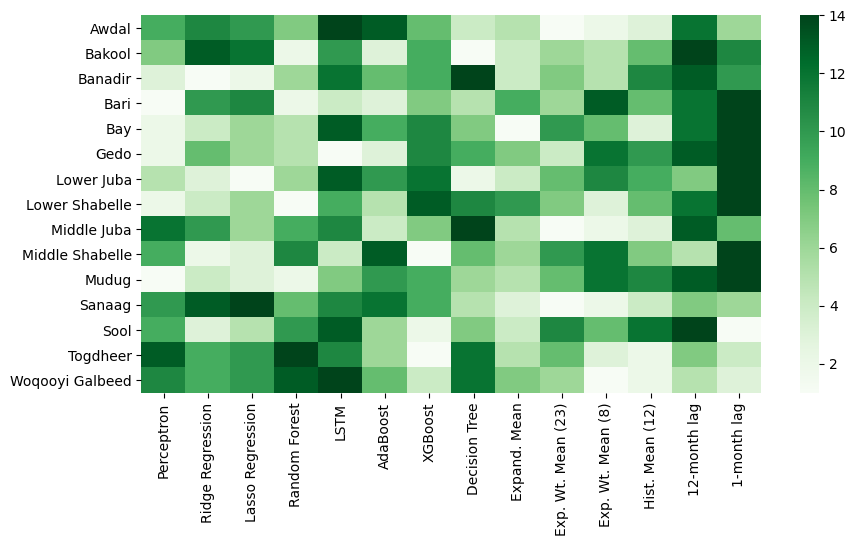}}
\caption{Ranked MSE}
\end{subfigure}
\caption{Models (x-axis) scored by region (y-axis) according to performance on the test dataset with a one-month horizon. As above, in order to ensure comparability across models, average MSE was calculated only over observations for which all models produced a prediction.  When examining performance in terms of raw MSE, we can see that regional differences in displacement magnitude dominate. When we rank model performance by region, however, we can see differences in performance between different classes of models. }\label{fig:rank}
\end{figure}

We ultimately chose not to transform the target variable (for example, by taking the log or dividing by the average number of arrivals per region) because discussions with field staff suggested that accurate prediction in high-volume regions \textit{should} be the priority. Instead, we adopted the proposed strategy of plotting ranked errors across all algorithms by region, which allows us to determine which algorithms are performing \textit{relatively} better or worse across a given region, and determine whether there is dataset-wide consensus on models across all regions. An example of such error rankings is shown in Figure \ref{fig:rank}. While it is hard to differentiate models when plotting raw MSE because regional differences in MSE are much greater than model-based differences in MSE, after ranking the models differences become clearer. For example, we can see that the expanding mean appears to dominate 1- and 12-month na\"ive lags in almost all cases. Machine learning models have mixed prediction quality, but the top models tend to do well in high-volume regions such as Banadir and poorly (relative to na\"ive benchmarks) in low-volume regions such as Awdal.

\subsection{{Model deployment and }next steps}
These predictive models are a work in progress and as such we are actively seeking to expand and improve the experiment prior to deployment. {Ultimately, we would like to promote model explainability and interpretability by building an interactive dashboard for use by humanitarian teams;
this would allow users to explore the predictions made by the different models described above and compare the forecasts of top performing models.}

 The fact that na\"ive benchmarks {currently} perform well implies that simple predictions made using historical arrivals data are competitive with the more complex machine learning predictions that incorporate detailed information on displacement drivers. In other words, average levels of past displacement are (fairly) good predictors for future displacement, at least from the perspective of our error metric (RMSE). {On the one hand, this is a function of the fact that we experimented with a much larger range of benchmarks than most comparable experiments, testing three different types of means -- expanding, exponentially weighted, and historical -- in addition to historical lags, and experimenting with different time horizons where applicable, for a total of 70 different benchmark models.\footnote{{ The most basic benchmarks -- 1 and 12-month lags -- do in fact perform poorly on both the train and test datasets.}} On the other hand,} this suggests that there is clearly room for improvement in the performance of the machine learning forecasts, {and one of our primary motives for developing this framework was to enumerate precisely which types of alternative design choices are possible. In ongoing work, we aim to improve performance by developing better infrastructure for running and evaluating experiments with these design choices, including different sets of input features, different transformations of the target variable, and different strategies for handling missing data.} 

{Despite their limitations, there are a number of reasons why we believe that machine learning models are valuable in this context, and have chosen to continue to develop these models.}  First, using average levels of past displacement to predict future displacement {(as in the na\"ive models)} will never generate unexpected predictions or flag unusual events; while machine learning models may fail to capture the magnitude of these events with precision, they do still have the potential to predict departures from existing trends, {and in particular sudden large changes or high frequency changes. In other words, na\"ive models can only ever capture na\"ive behaviors, whereas machine learning models can in principle capture much more complex relationships. As described above, some strategies to evaluate this possibility include developing classification models specifically targeted at detecting large increases or decreases relative to ``normal'', and/or adopting evaluation metrics that focus specifically on models' abilities to capture spikes or ``bursts''; these are areas for future work.} 

Second, given that machine learning models incorporate a wide variety of contextual variables to make predictions and that it is possible to quantify the importance of these variables in producing predictions, these models may be able to shed light on key drivers or correlates of displacement that can be examined via further analysis.  {This may be particularly important in displacement settings where many different contextual variables are available, but there are unclear or conflicting theories on which variables are most relevant and how exactly these variables affect the number of displaced people.}

{Third, there is a growing ecosystem of support for machine learning models and methods, and we expect that model performance and the available resources for modeling will continue to improve in the future; however, in policy settings these models are less commonly used than econometric models or ABM.  A natural extension of this work could directly compare the performance of machine learning models against these alternative models. Relative to such models, one advantage of machine learning is that it provides a structured, replicable approach to model construction and selection that focuses primarily on predictive performance; in contrast, economic/statistical forecasting models and ABM arguably allow more discretion for modelers to encode their assumptions about the world into the actual structure of the model (either through manual model specification in the case of economic/statistical forecasting models, or through the structure of interactions encoded into ABM). Another characteristic of machine learning models is that they can typically handle a larger number of feature variables -- including incorporating additional data sources such as satellite and social media data -- than econometric or statistical forecasting models, but do not require as detailed input data and assumptions as ABMs.}

\FloatBarrier
\section{Discussion and conclusions}
\label{sec:discussion}
The field of predictive analytics for humanitarian response is still at a nascent stage, but due to growing operational and policy interest we expect that it will expand substantially in the coming years. Although there has been a great deal of historical interest in early warning systems, some of the primary obstacles to the development of {predictive analytics and forecasting} systems have been the lack of available data and limitations in modeling and computational power. Recent years have seen substantial advances on both fronts.

To date, most research in this field comes from independent and situation-specific models. In order to further develop the field of predictive analytics for humanitarian response and translate research into operational responses at scale, we believe that it is necessary to better frame the problem and to develop a collective understanding of the available data sources, modeler decisions, and considerations for implementation. This paper has attempted to outline such a framework, drawing on examples from the literature and on our own experiences building predictive models. 
While there is a large vocabulary of readily-available predictive analytics techniques, we believe that this application setting -- the prediction of {refugee and IDP} flows -- poses unique challenges and considerations that complicate the use of standard models ``out of the box'', and have accordingly attempted to highlight many of these challenges and considerations above.

In the course of preparing this framework, we have found that relatively little is known about the structure of this prediction problem at the high level, and that there are a number of big-picture questions for which we lack empirical evidence. These include:

\begin{enumerate}
    \item \textit{How far in advance can we predict displacement?} There is a need to document how models perform as the time horizon extends into the future. Can we anticipate displaced flows one, six, or twelve months in advance? How does the performance of models degrade, {and uncertainty increase,} as predictions are made farther into the future? The forecast horizon will determine whether a model can be used for annual budgeting, short-term planning, early warning systems, or ``nowcasting'' emerging crises. 
    \item \textit{How much data is needed?} This question applies to both {the dimensions and the scale} of the prediction dataset. First: how much historical data is required to train a model, and at what point does adding additional historical data cease to meaningfully improve predictive accuracy? This will determine whether a model can be applied to new (rather than ongoing) crises, and whether it can be applied to crises for which historical data collection has been poor. Second: how does the introduction of additional features -- such  as variables extracted from social media or economic data -- influence predictive performance? This will determine the effort that should be spent on aggregating and integrating data from different sources. {Defining the right minimal data scale of the models that could be  generally applicable is key,} since it is difficult to find standardized {subnational} datasets that are available across countries.
    \item \textit{Do models generalize across borders {and contexts}?} ``Generalization'' can occur along two axes. First, can a model trained in one context be successfully applied to another? This will determine whether transfer learning can be used to address the ``cold start'' problem of predicting flows in newly emergent crises. Second, can a single model be successfully trained using data from multiple different displacement settings, and does {model performance} benefit from this attempt to generalize? Including input data from multiple settings means that more data will be available to the model, and that the model may be less likely to overfit to a specific scenario. However, it makes prediction more challenging because the model must learn about these disparate settings, and because the generative model of displacement may vary across different countries or regions.
\end{enumerate}
While we frame these questions as modeling challenges, they allude to deeper questions about the underlying nature of forced {displacement} which are of interest from a theoretical perspective. \textcite{schmeidl_early_1998} observe that there is often tension between academic and policy objectives, with academics focused on developing retrospective, higher-level insights while policymakers prefer forward-looking, ad-hoc approaches that can immediately inform decision making. By focusing on these high-level modeling challenges, we believe that these two perspectives can be unified: in addition to addressing the practical forecasting problem, answering these more substantive structural questions will contribute to the broader literature on forced displacement.

This is the paper we hoped we would find when we began to construct a predictive analytics model for Somali IDP flows. We hope that it is useful for others who are beginning a similar endeavor.

\newpage
\printbibliography

\FloatBarrier
\newpage

\appendix
\renewcommand\thefigure{\thesection.\arabic{figure}} 
\renewcommand\thetable{\thesection.\arabic{table}} 

\setcounter{figure}{0}  
\setcounter{table}{0}

%%%%%%%%%%%%%%%%%%%%%%%%%%%%%%%%%%%%%%%%%%%%%%%%%%%%%%%%%%%%%%%%%%
\section{{Modeling cards}}\label{sec:modeling_cards}
\begin{figure}[h]
\centerline{\includegraphics[width=4.25in]{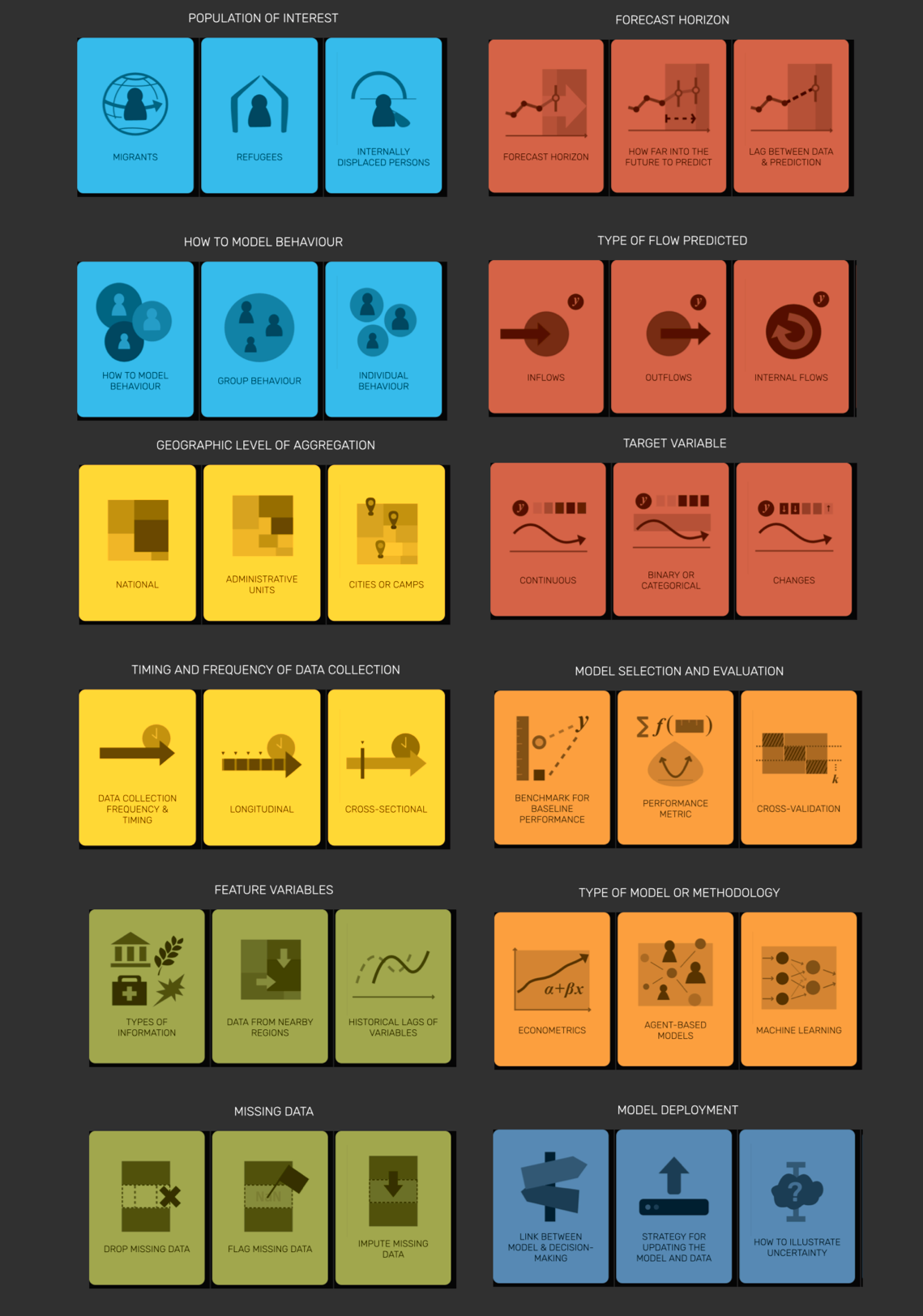}}\medskip
\caption{{A snapshot of interactive modeling cards, which were designed to illustrate this framework to practitioners.}}\label{fig:modeling_cards}
\end{figure}

\FloatBarrier
{In order to make the framework described in this article more accessible to practitioners, we have created an interactive visualization consisting of a set of ``modeling cards'', which allows practitioners to explore the different design choices available to them. Snapshots of these cards are shown in Figure \ref{fig:modeling_cards}, and a dynamic version of these cards is available at: \url{https://www.unglobalpulse.net/predictingdisplacement/}. }
  
%%%%%%%%%%%%%%%%%%%%%%%%%%%%%%%%%%%%%%%%%%%%%%%%%%%%%%%%%%%%%%%%%%
\FloatBarrier
\section{{Additional context}}
\FloatBarrier
Below, we provide context to frame the study of forced displacement. We {discuss} three categories of modeling needs, which are description, prediction, and simulation. We also discuss the formal terminology used to describe displaced persons, which informs their rights and the organizational mandate under which they fall. Finally, we discuss some of the characteristics of displacement flows, highlighting the diverse types of movements that are included in the term ``displacement''. 

\FloatBarrier
\subsection{A taxonomy of modeling needs}\label{sec:modeling_needs}
\FloatBarrier

While the nature of displacement flows is complex, the needs of humanitarian organizations are also plural and diverse. These organizations typically search for analytics tools that perform three key functions: description, prediction, and simulation. 

\textit{Description} refers to the most basic task of humanitarian data analysis: providing up-to-date, real-time information on the situation on the ground. Although there are tools and dashboards that measure displacement at different levels of geographic granularity, there is an interest in extending such systems to flag potential indicators of emergent crises; similar systems have already been developed for {climate and food security-related emergencies in Somalia \cite{fao_swalim_flood_2019, fsnau_early_2019}}.

\textit{Prediction} refers to the secondary task of estimating the magnitude of future refugee and IDP arrivals or needs. To be useful, predictions should meet a number of key criteria.  They should be \textit{timely}, since decision-makers must react quickly to changes on the ground; however, they must also look far enough into the future to give humanitarian teams the opportunity to respond. They should forecast an \textit{operationally useful quantity}, so that the predictions can actually be used in practice; for example, predictions may focus on arrivals rather than flows, since the ultimate preparedness goal is typically to anticipate how many people will enter a given region.    Predictive models should handle \textit{incomplete datasets}, since data does not always arrive on time for all variables and all regions, and it may be especially scarce in regions where violent conflict is high and humanitarian access to collect it is limited. Finally, predictive models should aspire to be \textit{trustworthy and explainable}. Humanitarian teams seek to understand why predictions have been made; they compare these predictions with their own priors, and can lose faith if the predictions are inaccurate or seemingly arbitrary.

\textit{Simulation} refers to the final and most complex task of modeling behavior in a way that generalizes to new or unseen situations. This capability is of strong interest to humanitarian teams. In particular, these teams may be interested in scenario forecasting based on different beliefs about future input variables (such as climate and conflict-related factors). Simulations can also be used to derive best- and worst-case scenarios by experimenting with different model assumptions.

\FloatBarrier
\subsection{The terminology of forced displacement}\label{sec:vocabulary}
\FloatBarrier

There are different classifications applied to people who move, depending on the factors that (presumably) drive their movement; formal definitions can be found in Figure \ref{fig:defs}. \textit{Migrant} is a general term that can refer to forcibly displaced people as well as those who are considered to have moved voluntarily, often for economic reasons; there is a theoretical debate between ``inclusivist'' and ``residualist'' definitions which contend that refugees should or should not be considered migrants, respectively \cite{carling_meaning_2021}. \textit{Refugees} are defined as those who have left their country due to persecution as a result of their nationality, {political beliefs, religion}, race, or membership of a social group. \textit{Internally displaced persons}, or IDPs, are defined as ``people who have been forced to flee their home but stay within their own country and remain under the protection of their government'' \cite{unhcr_global_2019}.
Together with other populations of interest such as asylum-seekers, returnees, and stateless persons, {refugees and IDPs} are collectively referred to as UNHCR's Populations of Concern \cite{unhcr_glossary_2021}.

While these terms may help to better describe population movements, they also have important practical implications. A person's classification as a migrant or refugee has legal consequences, influencing whether {they are} entitled to seek asylum and whether {they}  can be forcibly repatriated to their country of origin. This classification also has consequences for the mandates of humanitarian agencies.  For example, the debate about whether climate-driven displacement is forced or voluntary (i.e. whether one is a climate refugee or a climate migrant) has implications for whether UNHCR or IOM should be responsible for the emergency response. This classification also relates to notions of agency on behalf of the populations in question. Some scholars do not like the term ``forced'' displacement because they prefer to view displaced persons as autonomous and empowered individuals who make a \textit{choice} to respond to a difficult situation by leaving. Depending on the circumstances, migration may be a requirement for survival or a voluntary adaptation strategy \cite{idmc_no_2018}. 

\begin{table}[h]
\begin{tabular}{|lp{10cm}p{3cm}|}
\hline
    \textbf{Migrant}
    & ``An umbrella term, not defined under international law, reflecting the common lay understanding of a person who moves away from his or her place of usual residence, whether within a country or across an international border, temporarily or permanently, and for a variety of reasons.'' 
    & IOM Glossary on Migration \cite{iom_glossary_2019}
    \\
\hline
    \textbf{Refugee}
    &  Anyone who, ``owing to well-founded fear of being persecuted for reasons of race, religion, nationality, membership of a particular social group or political opinion, is outside the country of his nationality and is unable or, owing to such fear, is unwilling to avail himself of the protection of that country; or who, not having a nationality and being outside the country of his former habitual residence, is unable or, owing to such fear, is unwilling to return to it.'' & Convention and Protocol Relating to the Status of Refugees  \cite{unhcr_convention_2010}
    \\
\hline
    \textbf{IDP} 
    & ``Persons or groups of persons who have been forced or obliged to flee or to leave their homes or places of habitual residence, in particular as a result of or in order to avoid the effects of armed conflict, situations of generalized violence, violations of human rights or natural or human-made disasters, and who have not crossed an internationally recognized State border.''
    & OCHA Guiding Principles on Internal Displacement \cite{ocha_guiding_2001} \\
\hline
\end{tabular} \medskip
\caption{Official definitions of the terms migrants, refugees, and IDPs.}\label{fig:defs}
\end{table}

\FloatBarrier
\subsection{The characteristics of forced displacement}\label{sec:displacement_drivers}
\FloatBarrier

The classifications described above reflect underlying assumptions about the reasons why people move and the degree of choice involved in the movement. However, individuals rarely fit neatly into these definitions. In practice, people are displaced by a variety of driving factors, including: conflict, political instability or persecution, and economic or environmental conditions. It may be difficult to distinguish between these factors or to attribute displacement to a single cause; for example, environmental scarcity such as drought may generate violent conflict over resources such as water. Nevertheless, data collection practices rarely reflect this complexity: {for example, UNHCR's Protection and Return Monitoring Network (PRMN), which tracks displacement in Somalia, records only one reason for displacement.}

Attribution can be even harder when the displacement drivers are slow-onset in nature. For example, years of government repression can lead individuals to flee a country at the onset of a financial crisis, making it difficult to determine whether the displacement was political or economic in nature.  Researchers have attempted to characterize displacement drivers according to their time scale; for example, \textcite{schmeidl_early_1998} differentiate between \textit{root}, \textit{proximate}, and \textit{triggering} causes of displacement, which operate over the long term; over the past year; or over the past days or weeks, respectively. Similarly, \textcite{gemenne_why_2011} differentiates between gradual-onset and sudden-shock events. \textcite{gemenne_why_2011} also differentiates between \textit{temporary} or \textit{permanent} movement; \textit{national} or \textit{international} flows; and \textit{voluntary} or \textit{involuntary} departures.

%%%%%%%%%%%%%%%%%%%%%%%%%%%%%%%%%%%%%%%%%%%%%%%%%%%%%%%%%%%%%%%%%%

\FloatBarrier
\section{Supplementary tables and figures for the Somalia case study}\label{sec:somalia_details}
\FloatBarrier

\begin{table}[h]
\begin{tabular}{|p{2cm}|p{3cm}|p{8cm}l|}
\hline
\textbf{Data type} & \textbf{Source} & \textbf{Key variables used in analysis} & \textbf{Resolution} \\
\hline
Displacement & UNHCR PRMN \cite{unhcr_somalia_2019} & - Data on arrivals' previous and current regions & {Location-day} \\
\hline
Conflict & ACLED API \cite{raleigh_introducing_2010} & \makecell{  
			- \# incidents \\
			- \# fatalities }
			& Incident-level \\
\hline
Commodity prices & FSNAU price database \cite{fsnau_integrated_2019} & \makecell{  
		- The prices of wheat flour, water drums, local \\
		goats, red sorghum, petrol, charcoal, firewood, \\
		and imported red rice \\
		- The daily wage  \\
		- The Shilling-USD conversion rate}
		& Market-month \\
\hline
River levels & FSNAU EW-EA dashboard \cite{fsnau_early_2019} & 
		- The river levels recorded at the Baardheere, Belet Weyne, Buuale, Bulo Burto, Doolow, Jowhar, and Luuq stations 
		& {Station-month} \\
\hline
Environment and health & FSNAU EW-EA dashboard \cite{fsnau_early_2019} & \makecell{  
		- Rainfall  \\
		- Vegetation cover (NDVI)  \\
		- Cholera deaths  \\
		- Cases of cholera, measles, and malaria  \\
		- Hospital admissions for malnutrition (GAM) \\
		- Maize prices, the cost of the minimum basket, \\
		and terms of trade (goat-cereals, wage-cereals)} 
		& {District-month} \\
\hline
Distances & OSM \cite{open_street_map_overpass_2020} & 
		- Driving distances between region centroids 
		& District \\
\hline
\multicolumn{2}{|l|}{Manually calculated} & \makecell{
		- Whether pair of regions shares a border \\
		- Minimum direct distance between any two points in region pair \\
		- Dummy indicator for each region \\
		- Dummy indicator for each month of year (1-12) \\
		- A continuous variable counting months since January 2010 \\
		} & \\
\hline
\multicolumn{4}{p{17cm}}{\footnotesize 
ACLED = Armed Conflict Location \& Event Data Project; 
FSNAU = Food Security and Nutrition Analysis Unit, 
PRMN = Protection and Return Monitoring Network; 
NDVI = Normalized Difference Vegetation Index; 
EW-EA = Early Warning/Early Action; 
GAM = global acute malnutrition}
\end{tabular} 
\caption{Data sources and features used for the predictive model in Somalia.} \label{fig:data_sources}
\end{table}

%%%%%%%%%%%%%%%%%%%%%%%%%%%%%%%%%%%%%%%%%%%%%%%%%%%%%%%%%%%%%%%%%%
\begin{figure}
	\begin{subfigure}{.3\linewidth}
	\centerline{\includegraphics[height=1in]{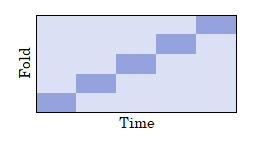}}
	\caption{Standard k-fold  (k=5)}\label{fig:tscv_a}
	\end{subfigure}
~
	\begin{subfigure}{.3\linewidth}
	\centerline{\includegraphics[height=1in]{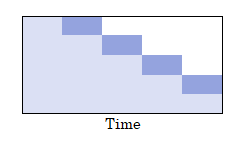}}
	\caption{Expanding window}\label{fig:tscv_b}
	\end{subfigure}
~	
	\begin{subfigure}{.4\linewidth}
	\centerline{\includegraphics[height=1in]{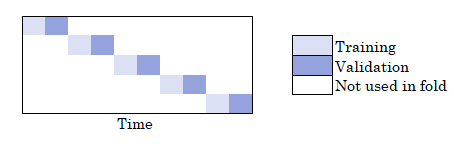}}
	\caption{Sliding window}\label{fig:tscv_c}
	\end{subfigure}
	
	\caption{Cross-validation strategies. {Adapted from \cite{packt_cross-validation_2019}}.}\label{fig:tscv}
\end{figure}

%%%%%%%%%%%%%%%%%%%%%%%%%%%%%%%%%%%%%%%%%%%%%%%%%%%%%%%%%%%%%%%%%%
\begin{table}
\centerline{\begin{tabular}{|l|rrr|r|}
\toprule
{} & Inflow poportion & Internal flow proportion & Outflow proportion & Proportion missing \\
\midrule
Awdal           &             0.37 &                     0.61 &               0.02 &               0.05 \\
Bakool          &             0.04 &                     0.59 &               0.37 &               0.09 \\
Banadir         &             0.72 &                     0.20 &               0.08 &               0.00 \\
Bari            &             0.28 &                     0.58 &               0.14 &               0.09 \\
Bay             &             0.14 &                     0.65 &               0.20 &               0.02 \\
Galgaduud       &             0.15 &                     0.66 &               0.19 &               0.44 \\
Gedo            &             0.16 &                     0.75 &               0.09 &               0.01 \\
Hiraan          &             0.03 &                     0.91 &               0.06 &               0.36 \\
Lower Juba      &             0.24 &                     0.67 &               0.10 &               0.00 \\
Lower Shabelle  &             0.04 &                     0.34 &               0.62 &               0.02 \\
Middle Juba     &             0.42 &                     0.32 &               0.25 &               0.20 \\
Middle Shabelle &             0.03 &                     0.84 &               0.13 &               0.19 \\
Mudug           &             0.16 &                     0.70 &               0.14 &               0.02 \\
Nugaal          &             0.45 &                     0.24 &               0.31 &               0.41 \\
Sanaag          &             0.13 &                     0.73 &               0.14 &               0.53 \\
Sool            &             0.16 &                     0.63 &               0.20 &               0.10 \\
Togdheer        &             0.19 &                     0.48 &               0.33 &               0.21 \\
Woqooyi Galbeed &             0.63 &                     0.27 &               0.10 &               0.03 \\
\bottomrule
\end{tabular}
}
\caption{The first three columns above show the proportion of inflows, outflows, and within-region displacement for each region in Somalia, aggregated over all observations in our sample. The last column shows the proportion of monthly arrivals counts that were missing by region in our dataset.}\label{fig:flows_and_missing}
\end{table}

%%%%%%%%%%%%%%%%%%%%%%%%%%%%%%%%%%%%%%%%%%%%%%%%%%%%%%%%%%%%%%%%%%

\begin{figure}
\centering
		\includegraphics[height=2.5in]{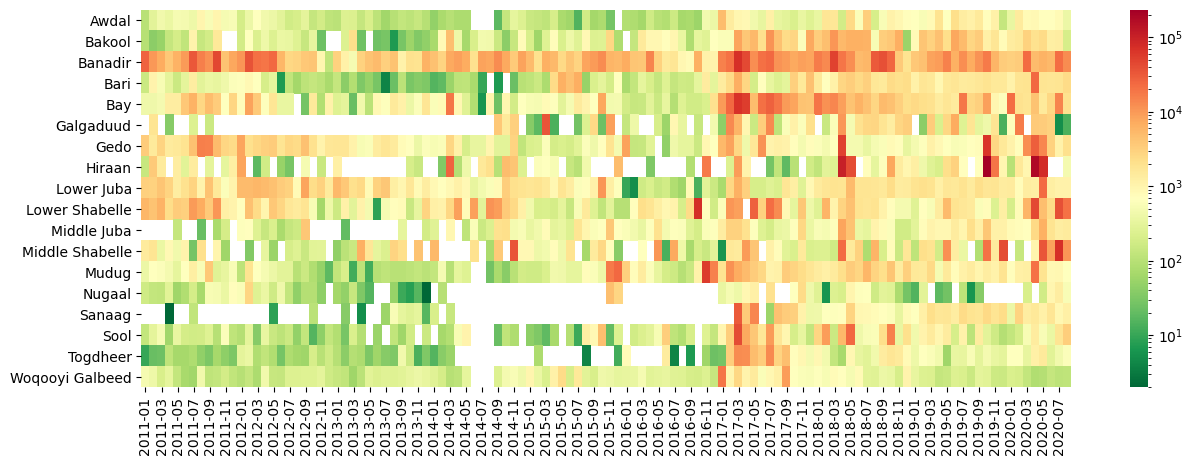}
		\caption{Monthly arrivals counts by region, 2011 - 2020.}
		\label{fig:arrivals}
\end{figure}

\begin{figure}
\centering
\includegraphics[height=5in]{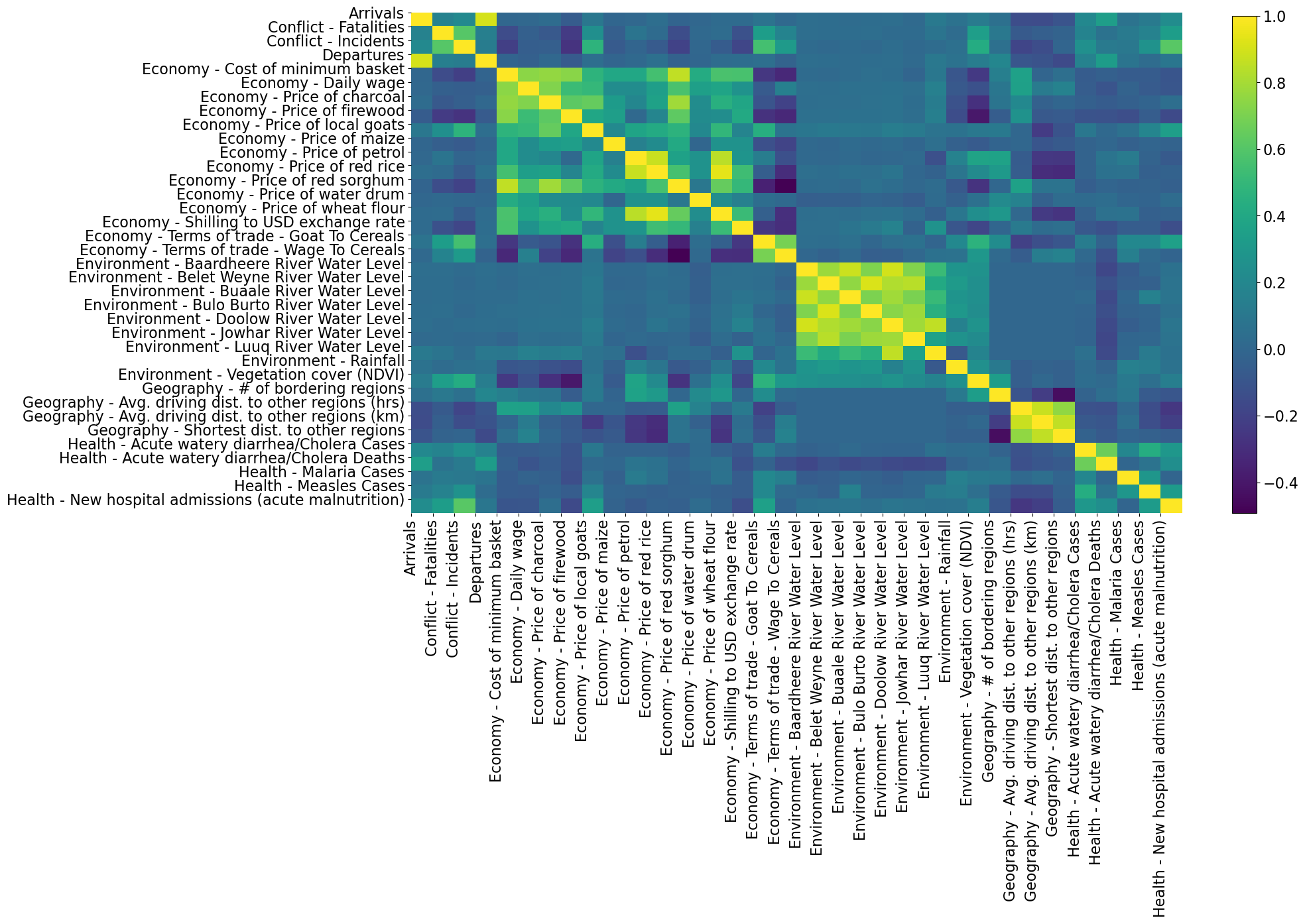}
		\caption{Plot of correlations between variables in the Somalia dataset.}
		\label{fig:corrplot}
\end{figure}

\end{document}